\documentclass[journal]{IEEEtran}

 \usepackage{subcaption}

\usepackage{lineno}
\usepackage{mathtools}
\usepackage{amsmath}
\usepackage{array}
\usepackage{mathrsfs}
\usepackage{amssymb}
\usepackage{booktabs}
\usepackage{algorithm}
\usepackage{algpseudocode}
\usepackage{multirow}
\usepackage{tabularx}
\usepackage{tabulary}
\usepackage{dcolumn}
\usepackage{threeparttable}
\usepackage{array}
\usepackage{bbm}
\usepackage{color}
\usepackage{soul}
\usepackage{graphicx}
\usepackage{wasysym}
\usepackage{amsmath}
\usepackage{amssymb}
\usepackage{amsthm}

\newcommand\myeq{\mathrel{\stackrel{\makebox[0pt]{\mbox{\normalfont\tiny iid}}}{\sim}}}

\begin{document}



\title{Modeling and Detection of Future Cyber-Enabled DSM Data Attacks using Supervised Learning}
%
%
%

\author{Kostas~Hatalis,~\IEEEmembership{Student~Member,~IEEE,}
        Parv~Venkitasubramaniam,~\IEEEmembership{Member,~IEEE,}
        and~Shalinee~Kishore,~\IEEEmembership{Member,~IEEE}
        
\thanks{K. Hatalis, P. Venkitasubramaniam and S. Kishore are with the Department
	of Electrical and Computer Engineering, Lehigh University, Bethlehem,
	PA, 18015, USA (e-mails: {\tt (kmh511,pav309,shk2)@lehigh.edu}).}
}

\maketitle

\begin{abstract}
Demand-side management (DSM) is a vital tool that can be used to ensure power system reliability and stability. In future smart grids, certain portions of a customers load usage could be under automatic control with a cyber-enabled DSM program which selectively schedules loads as a function of electricity prices to improve power balance and grid stability. In such a future, security of DSM cyberinfrastructure will be critical as advanced metering infrastructure, and communication systems are susceptible to hacking, cyber attacks. Such attacks, in the form of data injection, can manipulate customer load profiles and cause metering chaos and energy losses in the grid. These attacks are also exacerbated by the feedback mechanism between load management on the consumer side and dynamic price schemes by independent system operators. This work provides a novel methodology for modeling and simulating the nonlinear relationship between load management and real-time pricing. We then investigate the behavior of such feedback loop under intentional cyber attacks using our feedback model. We simulate and examine load-price data under different levels of DSM participation with three types of additive attacks: ramp, sudden, and point attacks. We applied change point and supervised learning methods for detection of DSM attacks. Results conclude that higher amounts of DSM participation can exacerbate attacks but also lead to better detection of such attacks, point attacks are the hardest to detect, and supervised learning methods produce results on par or better than sequential detectors.
\end{abstract}


\IEEEpeerreviewmaketitle

\section{Introduction}

\IEEEPARstart{D}{emand} Side Management (DSM) is an essential component in smart grids for planning, monitoring, and modification of consumer load levels. Furthermore, future cyber-enabled DSM will allow smart grids even higher levels of automated decision-making capabilities to selectively schedule loads on local grids to improve power balance and grid stability. Such a cyber approach relies heavily on real-time, two-way communication capabilities between a central controller and various adaptable loads. Research into security and reliability of the cyberinfrastructure that enables DSM is therefore vital. The main concerns in ensuring DSM security and safety lay in the feedback mechanism of real-time electricity pricing and distributed DSM controllable loads. Particularly in residential grids, each load contributes only a small amount of power and its compromising might not cause a noticeable impact on the power grid. However, a carefully planned or even chaotic cyber attack might impact other loads not under attack or not under DSM control by taking advantage of the feedback mechanism of load management. 

Two-way communication capabilities of advanced metering infrastructure (AMI) enables a utility or independent system operator (ISO), in the retail power markets, to collect high-resolution energy usage from consumers and enable dynamic pricing to adapt to consumer demand \cite{faruqui2010household}. Thus, AMIs provide an efficient way for ISOs to schedule prices and to then communicate those prices to consumers for automatic DSM control of certain portions of a load. AMIs can also provide practical ways for ISOs to set DSM goals such reducing peak or decreasing aggregate load levels through price influences. However, there are several vulnerabilities in AMIs that present noteworthy security issues since they are directly accessible by users \cite{grochocki2012ami}. Additionally, due to the large scale deployments of AMIs, ISOs encourage the utilization of marginally cheaper hardware which results in constrained computational resources to allow for robust security capacities, for example, intrusion monitoring. 

DSM programs utilize demand response, which is a specific tariff or program to motivate customers to respond to changes in price or availability of electricity over time by altering their regular electricity use habits. We take this a step further and look at future cyber-enabled DSM programs \cite{khan2016load} that will be able to autonomously control household loads such as water heaters and HVAC units based on RTP. As part of the reliable implementation of this future cyber-DSM, it is crucial to be able to understand the dependency between dynamic pricing and automatic demand response as well as the risks. We hypothesize that Cyber-DSM programs can be particularly vulnerable to cyber attacks such as false pricing information or direct load manipulation, especially when the participation rate in DSM is high. 

\begin{figure}[t]
	\centering
	\includegraphics[scale=0.9]{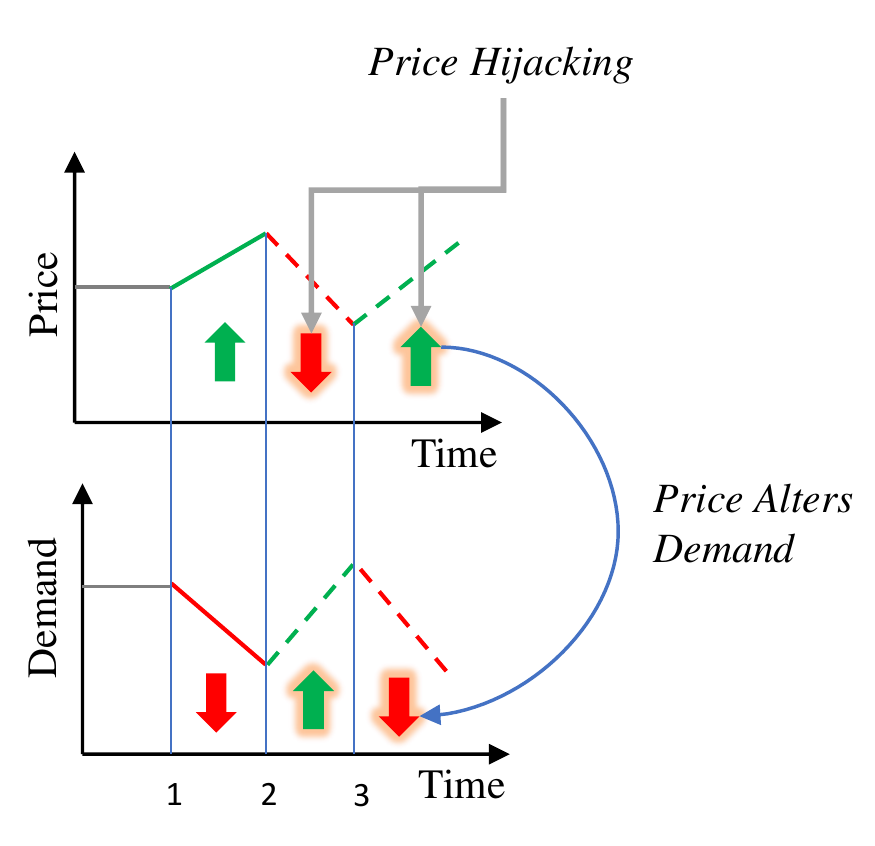}
	\caption{Feedback effect between price and DSM demand. As prices go up, demand decreases. But if prices are hijacked and false prices are fed to DSM systems then a false low price increases demand, and a false high price can decrease demand. The same is true if demand was altered by an attack. If load usage is increased by an attacker then prices would increase and vice versa.}
	\label{f:price_demand}
\end{figure}

Our work is thus motivated to study this vulnerability in DSM. We provide a mathematical formulation of the feedback between utilities and DSM systems, and then simulate, analyze, and test different detection methods for attacks on such feedback. This relationship between load and price is shown visually in Fig. \ref{f:price_demand}. As prices go up, demand naturally responds by decreasing. However, if AMIs are hijacked, and false lower prices are reported to DSM systems, then there will be an inappropriate increase in demand. A similar effect happens if an attacker directly controls user loads. Then a higher load usage by the attacker may inadvertently lead to higher prices for the rest of the grid. We present a standard of how attackers can exploit such a dependency. We propose a mathematical framework of the feedback between price setting and DSM systems to study how attacks can be structured and how to detect them. The main contributions of our approach can be summarized as follows:

\begin{enumerate}
	\item We formalizing the price to load relationship using an elastic demand model to achieve DSM goals.We simulate load and pricing data in a DSM system; in particular under a strategic conservation scheme as an example of a DSM goal.
	\item We propose two modes of attacks on DSM systems: false pricing data injection and direct load manipulation. We prove their equivalence and highlight three types of attacks that could be undertaken by each mode. We then empirically show how a high use of DSM can exacerbate attacks.
	\item We simulate these attacks and review sequential change-point and machine learning methods for detecting DSM attacks. In our main results, we demonstrate the impact of DSM on detection performance and identify what kind of detectors are effective for different attacks.
\end{enumerate}

In section \ref{s:review} we provide a literature review on DSM, and important DSM strategies real-time pricing and load forecasting. In section \ref{s:Microgrid} we apply the block bootstrap technique for simulating the non-DSM load distribution of a micro-grid of $N$ homes from template residential load time series. Dependency models for the feedback nature of load and prices are proposed in section \ref{s:model} where we showcase simulations of residential load and electricity prices when an automatic DSM program controls certain portions of consumer demand as a function of price. In section \ref{s:attack} we present two modes of cyber attacks, direct load manipulation attacks and price data injection attacks that can have a significant influence on the feedback of load and price. We prove these two attacks are equivalent. We conclude in section \ref{s:future} with possible directions of future work.

\section{Background} \label{s:review}

There are multiple strategies to accomplish DSM like load forecasting and real-time pricing which are used for load management on in-home energy management systems. So before introducing our models on the relationship between price setting and demand response by cyber DSM systems, it is crucial to review foundational material on DSM. In this section, we provide a brief overview of DSM goals and approaches, real-time pricing, and load forecasting. We also review different pricing simulation schemes which will play a role in modeling RTP.

\subsection{Demand Side Management}

\begin{figure}[t]
	\centering
	\includegraphics[scale=0.18]{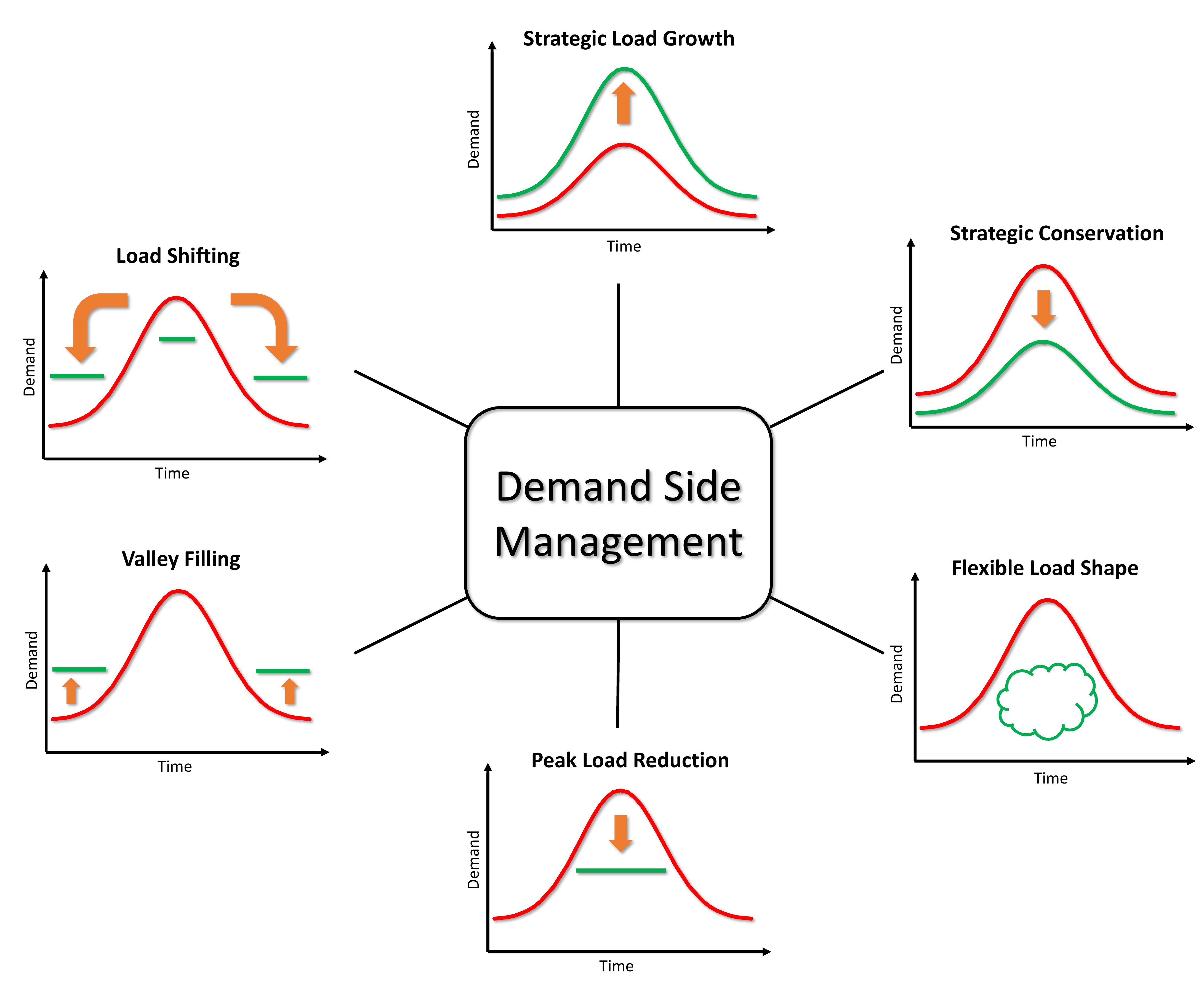}
	\caption{Various demand side management goals.}
	\label{f:strategic}
\end{figure}

DSM is an active and voluntary approach for reducing electricity use through activities or programs that promote electric energy efficiency, conservation, or more efficient management of electric energy loads \cite{palensky2011demand}. Very often financial incentives and educational programs are used to modify consumer demand. More specifically, the main goals of DSM are peak clipping, valley filling, load shifting, strategic load growth, flexible load shaping, and strategic conservation. These goals are summarized in Fig. \ref{f:strategic}. In these goals, consumers are encouraged to use less energy during peak hours, or to move the time of energy use to off-peak times such as nighttime, or reduce overall consumption. Other applications for DSM is to aid grid operators in balancing intermittent generation from wind and solar farms due to their volatility nature which may not coincide with energy demand at different times of the day.

In our study, we focus on modeling and simulating the DSM goal of strategic conservation, due to its simplicity and essential use in the smart grid. This goal also makes it easier to study attacks on DSM by modeling strategic conservation as a general reduction in load. Attacks then could stand out more versus goals like flexible load shaping. More specifically, this DSM goal aims at reducing aggregate load demand through directed reduction of electricity consumption. The successful implementation of strategic conservation programs usually requires some combination of financial incentives to customers, the promotion of energy-efficient building standards, and appliance efficiency improvement. Strategic conservation also requires a more excellent knowledge on the part of the utility concerning customer behavior. We envision in the future that AMI and smart appliances in residential DSM programs will automatically control specific portions of consumer load as a function of real-time electricity prices to achieve the goal of strategic conservation.

Most DSM programs are formulated as an optimization problem as follows
\begin{equation*}
	\min_{P_t} \sum_{t=1}^{T} (L_t - L_{t}')^2,
\end{equation*}

\noindent where $P_t$ is RTP at time $t$, $L_t$ is actual load, and $L_{t}'$ is the target load level the ISO is interested in achieving via DSM. The aim is to choose a price $P_t$ for each time step such that the actual load would reach as close as possible to the target level. In \cite{logenthiran2012demand}, a DSM strategy was proposed based on heuristic optimization to shape the load curve close to the desired shape. A  heuristic-based evolutionary algorithm was used to solve the above minimization problem. A multi-agent game theoretical DSM approach is proposed in \cite{mohsenian2010autonomous}. The authors use game theory and formulate an energy consumption scheduling game, where the players are the users, and their strategies are the daily schedules of their household appliances and loads. In \cite{gelazanskas2014demand}, the minimization problem is solved by utilizing a feedforward neural network to map the nonlinear relationship between price and load. Recently, the DSM problem is addressed in \cite{batista2018demand} as a multi-objective optimization problem that also seeks to balance other merit functions such as energy production cost, costumer’s preferences, and other constraints. 

\subsection{Load Forecasting} \label{s:load_forecasting}

Load forecasting (LF) techniques are an essential component for RTP and other ISO operations by predicting future energy requirements of a system from previous data and weather conditions. It is recognized as the initial building block of utility planning efforts and ensures the balance between supply and demand of energy, LF thus plays a vital role in our DSM formulations. For a given system and requirements, LF provides predictions for specific periods. These periods are divided into short, medium, and long term forecasts. Short term LF is used to predict load on an hourly basis up to 1 week for daily operations and cost minimization. Medium-term LF typically predicts load on weekly, monthly, or yearly basis for efficient operational preparations. Long term LF is used to predict load up decades ahead to facilitate grid and generation expansion planning. In this work, we look at short term LF on the resolution of one hour up to a week.

LF models can be divided into two approaches \cite{khan2016load}, the first being statistical based modeling and the second being machine learning. The statistical approach can be further broken down into regression and time series models. Multiple linear regression can be used with the weighted least squares estimation technique to form a relationship between different independent covariates that load depends on such as weather conditions. Regression models have been applied in LF in different works such as in \cite{dudek2016pattern}. Time series models are also prevalent to apply to LF. The most common model is the autoregressive moving average (ARMA) model and its variants that include components such as integration (I), fractional integration (FI), multivariate series (V), seasonality (S), exogenous (X) data, conditional heteroskedasticity (CH), and nonlinearity (N). For our formulation, we employ SARIMA for LF.

Hyperparameters of ARMA models can be solved using Box-Jenkins decomposition or grid search with an Akaike information criterion. Various studies have looked at all the different ARMA models for LF \cite{pappas2010electricity}. Other time series methods for LF include simple exponential smoothing \cite{christiaanse1971short} and the Holt-Winters seasonal method \cite{taylor2010triple}. Time series analysis and regression analysis share many models and ideas, but they are theoretically different. Time series analysis first deals with time indexed stationary data and account for the autocorrelation between time events. In regression we assume there is no autocorrelation, and that all observations are independent and identically distributed. Furthermore, we also assume in regression the data is homeostatic and does not exhibit multicollinearity.

Most recently, machine learning methods have seen a huge spike in LF research. Machine learning models are data-driven, typically providing a nonlinear fit to input covariate data to predict load. Advantages of this approach include not needing preconditions for data such as stationarity (a requirement for most time series methods), excels at modeling nonlinear dependences, and can fit large data sets. Disadvantages for most machine learning models are that most hyperparameters are continuous (difficult to tune), they require extensive feature engineer, and may get stuck in local minimums. Models for LF include support vector machines \cite{kavousi2014new}, feedforward neural networks \cite{baliyan2015review}, recurrent neural networks \cite{kong2017short}, random forests \cite{dudek2015short}, and ensemble learning \cite{li2016ensemble}.       

In all the various LF approaches, benchmarks are required to compare the prediction performance of our models. The most common benchmarks are the naive persistence and seasonal methods. In the naive persistence method a prediction of the load for time $t$ is equal to the value from the previous time step $t-1$ and in the seasonal persistence method a prediction of the load for time $t$ is equal the same hourly value from the previous day $t-24$.

\subsection{Real Time Pricing}

Every consumer of electricity is charged with a certain amount per kilowatt hour (kWh) of energy. Such a charge is done to cover the costs associated with the generation, transmission, and distribution of electricity. The two main types of costs are operational and fixed costs. During the 20th century, tariffs have been used to recover costs. Lately, clever pricing schemes have been developed to meet the requirements of modern power systems \cite{chakraborty2014smart}, such as, real-time pricing (RTP) where consumers are charged with a price nearest to the real price of generation at a specific interval in time. RTP plays an integral part in time-based DSM programs that makes consumers choose the time of consumption of power as a response to prices \cite{nazar2012time}. Cyber-enabled DSM programs are an automated form of time-based DSM programs. 

There are two types of RTP schemes, hourly pricing and day ahead pricing. In the first type, the price of electricity is released on an hourly basis for the next hour. In day ahead pricing, prices are announced for the next 24 hours based on predicting the load demand and the cost of generation. RTP signals combined with DSM automation at the consumer level provides benefits to both consumers and utility. A properly designed RTP scheme increases the reliability of the grid, reduces associated costs with generation, and lowers electricity bills of consumers. Further review of RTP and other dynamic pricing schemes can be found in \cite{khan2016load}.

\section{Microgrid Simulation}  \label{s:Microgrid}

Before modeling the load-price dependency of a DSM system, we need first to obtain some ground truth data of what load data from a residential micro-grid looks like without the presence of DSM, where we assume the elasticity of demand to price is very low. To do this, we use the power time series from several homes as templates for our grid and then generate artificial $ N $ household datasets. There are several ways to generate artificially residential load data, such as by using power grid simulators such as MATPOWER \cite{zimmerman1997matpower} or GridLAB-D \cite{chassin2008gridlab}. Our object is to create a model free real-data driven alternative to grid simulators. We are able to generate unlimited but plausible univariate load data to serve as the base demand for sample households before the application of a DSM system. 

\subsection{Data Source}

We use the UMass Smart* \cite{barker2012smart} dataset, 2017 release, for the simulation of micro-grid load time series. The Smart* project built a data collection infrastructure that records data from a variety of sensors deployed in real homes. Their infrastructure supports both pulling data by querying individual sensors and pushing data from sensors to a gateway server, which ran on their software tools. The 2017 Home dataset release is comprised of electrical power readings from seven homes from 2014 to 2016 at a minute resolution. It includes readings from individual appliance sensors as well as total power usage of each home. We chose to use these seven datasets as template homes in simulating the power usage of a micro-grid due to the breadth of the data collected. For DSM attack research, these datasets can help model an attacker compromising individual appliances. 
For each home the power consumption is given in kW for every minute. We convert this time series to kWh with a resolution of one hour which is common for smart meter readings and real time price modeling \cite{khan2016load}. We do this by obtaining the average power consumption within an hour and multiplying it by the time period as such
\begin{equation}
E_{(kWh)} = t_{(hr)} \times \frac{1}{60} \sum_{i=1}^{60} P_{(kW)}^{i}
\end{equation}

\subsection{Block Bootstrap Simulation} \label{s:bootstrap}

In the generation of new time series from sample data, several approaches can be applied depending on the statistical properties of the series. Data that is stationary can be modeled and generated using an ARMA process \cite{brockwell2002introduction}. An ARMA model is fitted to the data, and then future data is sampled from the ARMA distribution. If there is no serial correlation, then the distribution of some sample data can be modeled using Markov Chain Monte Carlo \cite{gamerman2006markov}, and new data can be sampled from this estimated distribution. However, in the case that data exhibited autocorrelation and non-stationarity in the presence of a periodic seasonal pattern, then a natural choice is to use the block bootstrap method \cite{politis2004automatic}.

\begin{figure}[t]
	\centering
	\includegraphics[scale=0.26]{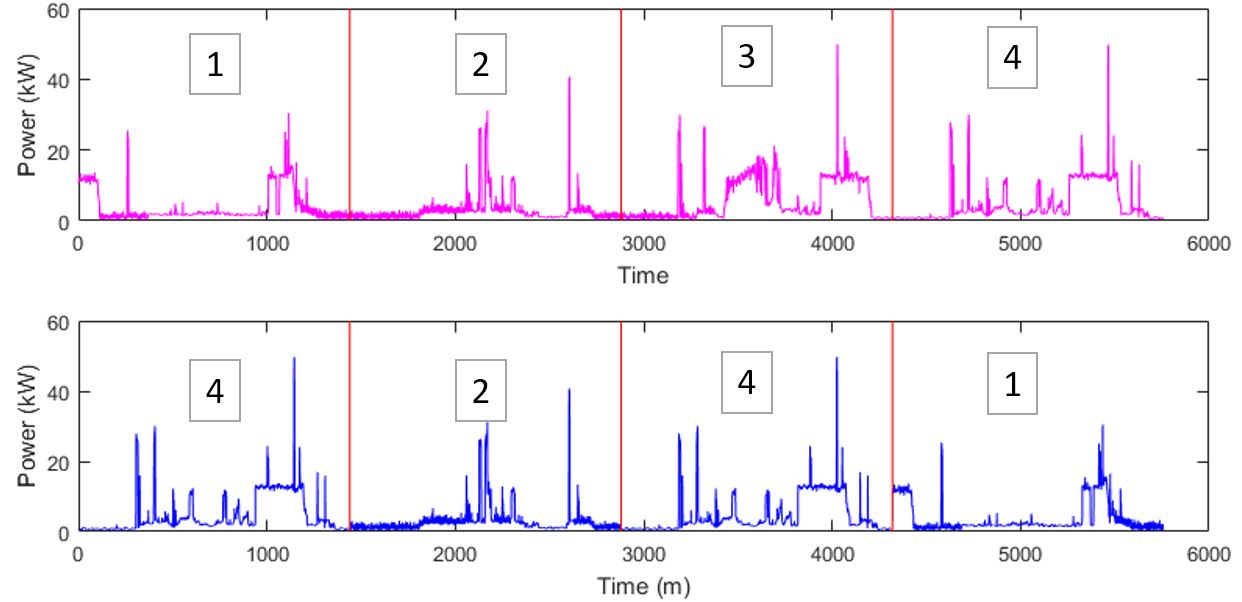}
	\caption{Process of block bootstrap simulation of a new home power usage (bottom) from a template home (top). Example simulation samples are taken from four days from the template series with replacement.}
	\label{f:bootstrap}
\end{figure}

Bootstrap is used in simulation statistics for estimating the distribution of a statistic such as mean or variance. This is particularly useful when there is no analytical form to estimate the density of our underlying statistics. A bootstrap analysis is conducted by using the Monte Carlo algorithms with replacement. Data is sampled with replacement until a new set is formed and then statistics are calculated from that new set. The process can be repeated to get a more precise estimate of the Bootstrap distribution and to form confidence intervals for those statistics. The block bootstrap is used when the data, or the errors in a model, are correlated. The block bootstrap attempts to replicate the serial correlation by resampling blocks of data instead of individual observations. This is why the block bootstrap is used primarily with correlated time series. In block bootstrap, blocks sampled can overlap or be non-overlapping. For load time series simulation we use block bootstrap with non-overlapping blocks to preserve the daily seasonal pattern of power consumption. The process of block bootstrap simulation of a new home power usage from a template home is shown in Fig. \ref{f:bootstrap}.

\section{Dependency Model} \label{s:model}

\subsection{Modeling Elastic Demand}

For analyzing the feedback dependency between load and price in a DSM setting it is first required to define a supply and demand relationship of electricity. To do so we utilize the well known measure in economics, the Price Elasticity of Demand (PED) \cite{thimmapuram2010modeling} which can be given by
\begin{equation} \label{e:PED}
\epsilon_d = \frac{dL}{dP} \cdot \frac{P}{L}.
\end{equation}

\begin{figure}[t]
	\centering
	\includegraphics[scale=0.6]{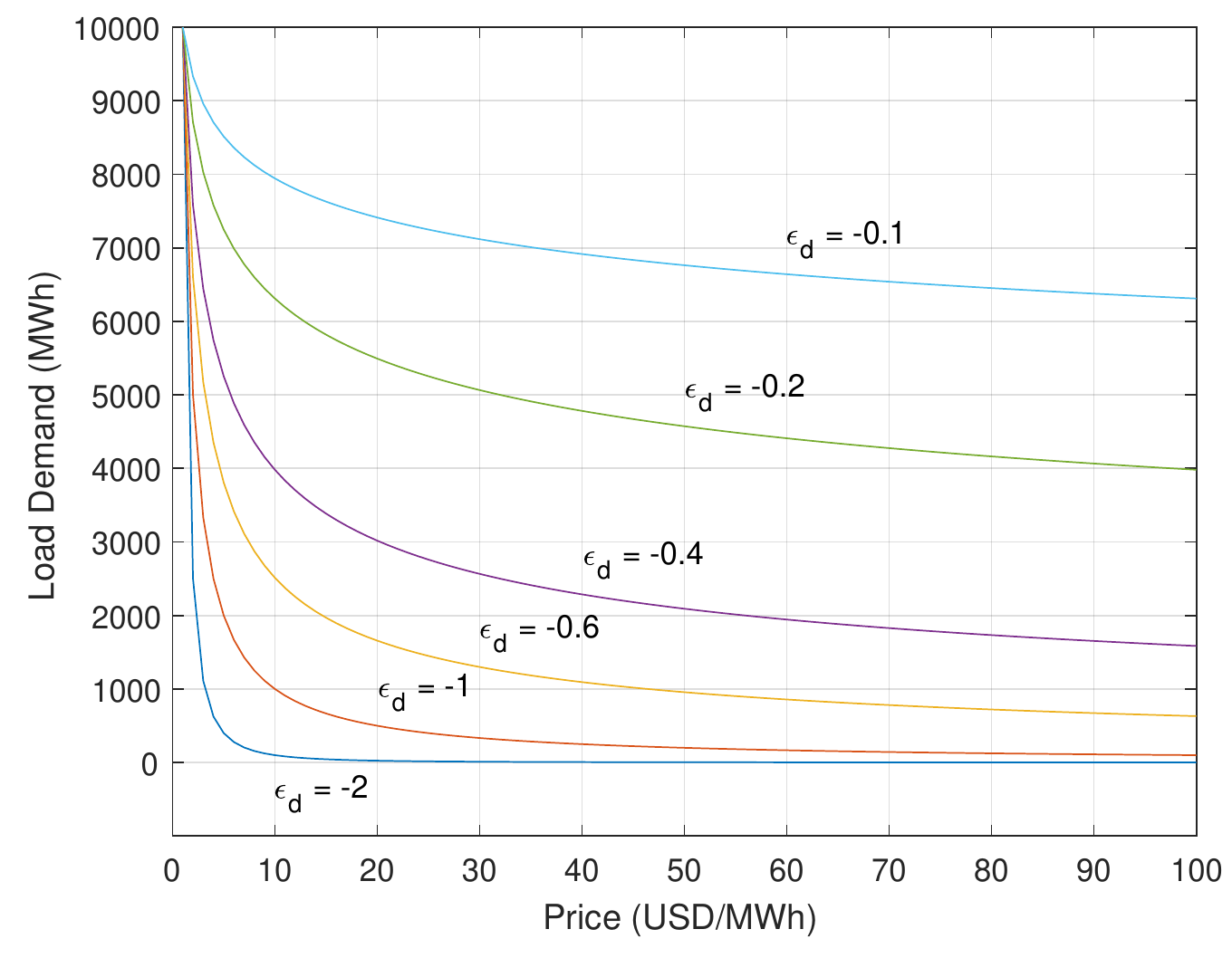}
	\caption{Load as a function of price with arbitrary price range \$1-100, $ \alpha=10,000 $, and $ \epsilon_d=$ -0.1,-0.2,-0.4,-0.6,-1, -2.}
	\label{f:demand_curve}
\end{figure}

\noindent PED shows the responsiveness of the Load (L) demanded of electricity to a change in its Price (P). An absolute value of PED = 1 shows unitary elasticity.  For instance, when  $\epsilon_d = -1$ then a 1\% change in the price will have a 1\% change in the load demanded. As prices increase, load will decrease. When absolute PED falls between 0 and 1, this signifies that the demand for load is inelastic, while a value greater then 1 says that the demand is elastic. When   $|\epsilon_d| = 0$ the demand is perfectly inelastic. A change in price has no affect on the load. While  $\epsilon_d = \infty $ represents perfect elasticity. If $\epsilon_d$ is constant the whole demand curve then,
\begin{equation} \label{e:L}
\begin{split}
\frac{1}{L} dL &= \epsilon_d \frac{1}{P} dP \\
\int \frac{1}{L} dL &= \epsilon_d \int  \frac{1}{P} dP \\
\ln (L) &= \epsilon_d \ln (P) + c \\
e^{\ln (L)} &=e^{\epsilon_d \ln (P) + c}\\
L &= e^{\epsilon_d \ln (P)} \cdot e^{c}\\
L &= \left( e^{\ln (P)} \right)^{\epsilon_d } \cdot e^{c}\\
L &= P^{\epsilon_d}\cdot e^{c}\\
\text{substitute } a &= e^{c} \\
\Rightarrow L &= a \cdot P^{\epsilon_d} 
\end{split}
\end{equation}

\noindent where $a$ is a scaling constant. An example demand curve estimated from Eq. \ref{e:L} can be seen in Fig. \ref{f:demand_curve}. The figure also showcases the nonlinear relationship between load and price where as the price, the independent variable, increases the load demanded, the dependent variable, decreases. 

\subsection{Modeling Consumer DSM}

In modeling the relationship between load and price under a DSM program, it is important to define the individual loads of each customer in determining aggregate load. For a customer who does not participate in a DSM program their load is determined by the stochastic demand process of users actions such as watching tv, using the AC, etc. Demand is impacted by multiple factors such as user preferences, weather, and time of day. In this process, electricity prices have a small influence on demand - individual customer demand is fairly inelastic to price. Following the derivation in Eq. \ref{e:L} we define the load usage of an individual customer $i$ for time $t$ as 
\begin{equation*}
\phi_{t,i} = \theta_{t,i} (P_{t}+P_c)^{\epsilon^{d}_{t,i}},
\end{equation*}

\noindent where $ P_{t} $ is the RTP for time $t$, $P_c$ is constant of the retailer's market costs which does not vary with RTP, $ \theta_{t,i} \geq 0 $ is a scaling factor representing the stochastic process that determines the user load, and $ \epsilon^{d}_{t,i} $ is the elasticity coefficient for the individual customers sensitivity to price changes. It can vary over time but without DSM incentives most users have a fairly inelastic PED. 
For experimental purposes, in modeling individual user loads we set $ \phi_{t,i} $ equal to simulated bootstrapped user load profiles defined in Section \ref{s:bootstrap}. We assume that prices, user preferences $ \theta_{t,i} $, and $ \epsilon^{d}_{t,i} $ have been absorbed in the calculation of the simulated load series. Thus we use $ \phi_{t,i} $ as a reference point to how much electricity  a user wants to consume without the influence of a DSM program. We model the task of the ISO as modifying $ \phi_{t,i} $ to some desired load levels.

The goal of DSM is to motivate the consumer to use less energy typically during peak hours. A DSM program could have a number of goals all aimed at reducing load usage. These may be peak clipping where peak load is reduced, load shifting where times of higher load are reduced and times of low load are increased, and many more strategies. There are multiple ways a DSM program can achieve these goals such as financial incentives, education, and regulation. In this work we look at the near term future where in a home with a smart meter and smart appliances, users can participate in a DSM program that automatically adjusts the load of interconnected appliances (such as the water heater) as a function of price and a DSM elasticity term that determines how fast or slow a users load under DSM control. 

Realistically, only a certain portion $ \kappa_{t,i} \in [0,1) $ of customer $i$'s power usage will be under control of a cyber DSM program. There will always be some stochastic component of power usage such as using a microwave oven or electric hairdryer. An ISO can signal a users DSM program by setting prices where smart appliances adopt to price changes with certain elasticity. We do no model direct load control. If prices are set too high it will signal the DSM component of a users demand to start using less load. If prices are set low then it signals the DSM program to increase load usage. We model this DSM component as such 
\begin{equation} \label{e:user}
l_{t,i} = (\kappa_{t,i} \phi_{t,i}) P_{t}^{\epsilon^{dsm}_{t,i}} + (1-\kappa_{t,i}) \phi_{t,i} ,
\end{equation}

\noindent where the first part $ (\kappa_{t,i} \phi_{t,i}) P_{t}^{\epsilon^{dsm}_{t,i}} $ is the load level customer $i$ allows the DSM program to determine as a function of price $ P_{t}$ and the DSM's elasticity to price $\epsilon^{dsm}_{t,i}$. Price elasticity of DSM $\epsilon^{dsm}_{t,i}$ may very over time and customer and affects how much power usage should be affected by price. The second portion $ (1-\kappa_{t,i}) \phi_{t,i} $ of a customers load is the stochastic component. The DSM portion parameter $\kappa_{t,i}$ is defined as a function of time where users can add or remove house loads under DSM control over time.  The term $\kappa_{t,i}$ can be modeled as a random variable (e.g. Uniform or Gaussian) or as a fixed constant for users. Once each individual customers load has been determined then total load (modified by a cyber-DSM program) for time $t$ for $N$ customers is calculated by 
\begin{equation} \label{e:L_total}
L_t = \sum_{i=1}^{N} l_{t,i}.
\end{equation}

\noindent We also define the aggregate base load as 
\begin{equation} \label{e:Phi}
\Phi_t = \sum_{i=1}^{N} \phi_{t,i},
\end{equation}

\noindent which represents the total demand had there not been a DSM program for a time period $t$ (ie $ \kappa_{t,i}=0, \forall_i $).

\subsection{Modeling DSM Goals}

The ISO has different DSM goals for reducing a customers load profile as mentioned before these goals can be to reduce peak load or increase load in periods of low demand. Two ways for a DSM program to control load are to make it a function of electricity prices or for an ISO to directly control a homes energy usage by programming its smart appliance. The effects of this feedback control approach can also be seen in Fig. \ref{f:price_demand}. Given the privacy issues of having an ISO directly control a homes energy usage, we model the indirect approach of achieving DSM goals through setting electricity prices. 

We introduce an approach how an ISO can set RTP, on an hourly basis, as a function of aggregate load to achieve a desired load level $L'_{t}$. The approach takes as input a forecast $\hat{\Phi}_{t}$ of the base aggregate load to calculate price $P_t$. This prediction can be defined as 
\begin{equation} \label{e:Phi_prediction}
\hat{\Phi}_t = f_{pred} \left( \Phi_{t-k:t-1}, P_{t-k:t-1}, X \right), 
\end{equation}

\noindent where inputs to the prediction model are past base load $\Phi$ and price $P$ values from time $ t-k $ to $ t-1 $, and other predictor variables $X$ such as time-of-day and weather information. Various types of prediction models can be used for $ f_{pred} $ such as neural networks as reviewed in section \ref{s:load_forecasting}. We now define RTP based on the formulation in Eq. \ref{e:L} as such
\begin{equation} \label{e:P2}
\begin{split}
P_t &= \left( \frac{L^{*}}{\hat{\Phi}_{t}} \right)^{1/\hat{\epsilon}^{dsm}}
\end{split}.
\end{equation}

\noindent Where,
\begin{equation} \label{e:L*}
\begin{split}
\text{Goal 1: } L^{*} &= L'_{t} \\
\text{Goal 2: } L^{*} &= L'_{t} +(L'_{t-1}-L_{t-1})
\end{split}.
\end{equation}

\noindent The component $ L^{*} $ adjusts the RTP based on two goals the ISO may have. The first goal is to adjust the price to push power usage directly to the target level $ L'_{t} $ with the assumption that there is near 100\% DSM participation by all customers. The expectation is that if demand for time $t$ is $ \hat{\Phi}_{t} $, then a price point is set to push load usage to $ L'_{t} $. Of course,if participation is less then 100\%, which is more likely, then the target level $ L'_{t} $ will not be reached by $P_t$. So, to push the aggregate load, from all users, as close as possible to the target load level, with an unknown amount of participation, then a penalty would need to be added to $P_t$. We model this as Goal 2 where the idea is to affect the power usage of those under cyber-DSM control even more than goal 1 to compensate for users who are not participating in DSM. 

Some users will not be participating, or only have a small portion of their power usage under control by the cyber-DSM program. We model their remaining power usage as inelastic to RTP. Thus, to push aggregate load to a target level, taking into account some load usage is inelastic, we need to push RTP much higher or lower to have a bigger effect in pushing DSM controlled load closer to the target load. This is what Goal 2 attempts to do, with the component $ L^{*}= L'_{t}+(L'_{t-1}-L_{t-1}) $ taking the target load level for time $t$ and adding the difference from the previous target load $ L'_{t-1} $ and realized load $ L_{t-1} $ as a penalty to adjust RTP to compensate for the difference. If $ L^{*} < 0$ then we set $ L^{*} = 10$ or to some arbitrary small target value. By subtracting the difference between the previous load and target level, we make up for users not participating in DSM by forcing DSM users a higher price to push their load even lower.

The term $\hat{\epsilon}^{dsm}$ in Eq. \ref{e:P2} is an estimate of the price elasticity of DSM of the whole grid; if individual user coefficients $\epsilon^{dsm}_{t,i}$ are unknown then $\hat{\epsilon}^{dsm}$ can be estimated from observing past values of price and load under different levels of DSM control. Alternatively, the ISO can define $\epsilon^{dsm}$ for all household cyber DSM programs. The formulation in Eq. \ref{e:P2} sets prices by comparing the adjusted target load for time $t$ to the forecasted base demand $ \hat{\Phi}_{t} $ for the same time. This demand $ \hat{\Phi}_{t} $ would be the level if no load was under DSM influence, thus to influence and alter it, $ \Phi_{t} $ needs to be estimated as accurate as possible.

\begin{figure}[t]
	\centering
	\includegraphics[scale=0.5]{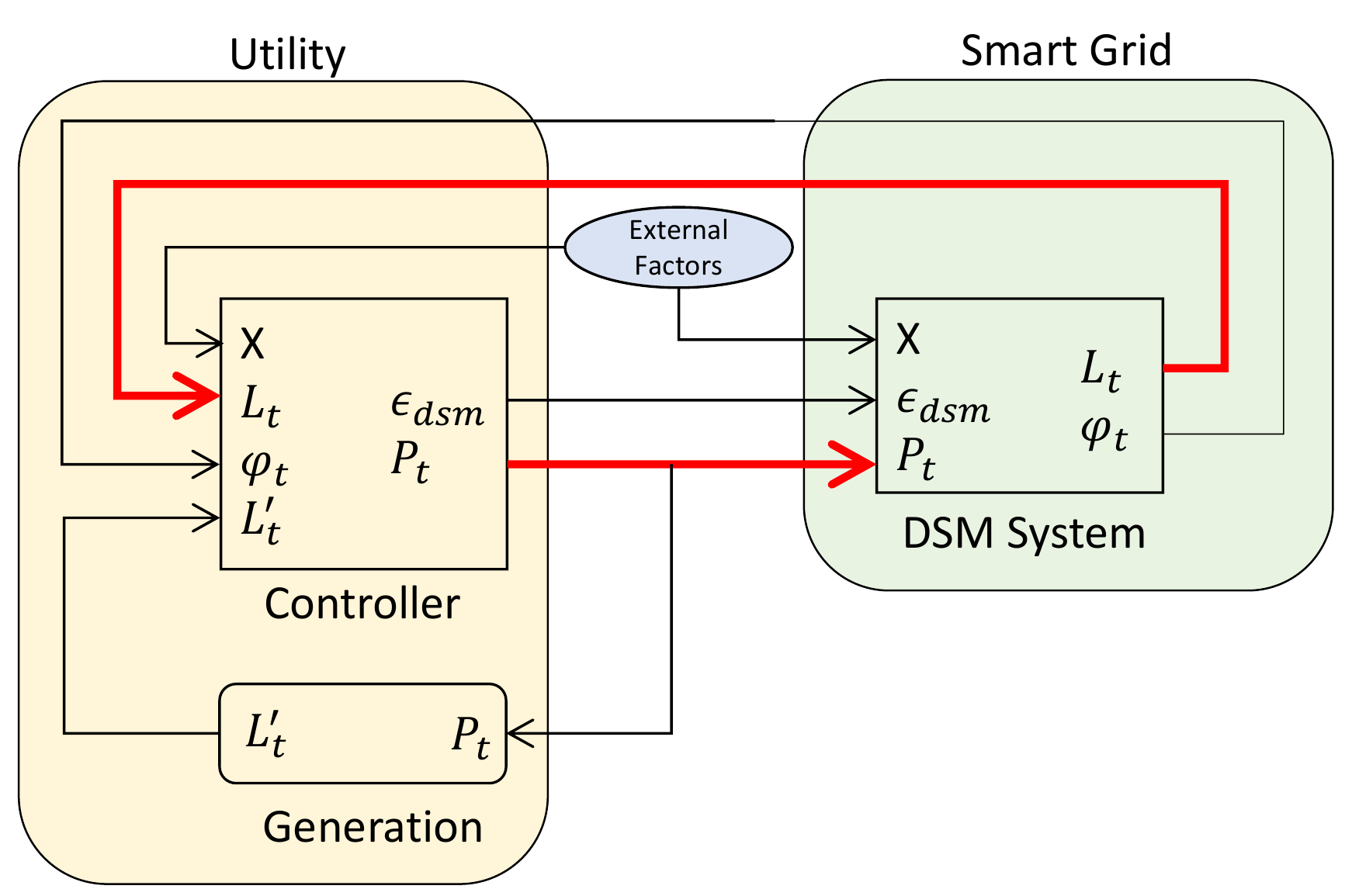}
	\caption{Block diagram highlighting the feedback between the utility and grid.}
	\label{f:diagram}
\end{figure}

In our approach, if the aggregate load is above the target load, RTP is set higher to decrease demand. If the aggregate load is lower than the target load, then the price is decreased to increase demand. Thus, as also can be observed, in Eq. \ref{e:P2}, there is direct feedback between price and load Eq. \ref{e:user}. The block diagram in Fig. \ref{f:diagram}  also outlines this feedback that showcases the relationship between the utility and grid. Generation sets the target load based, on the price and supply of power, and the controller sets the price signal and the elasticity of demand coefficient for DSM systems.  The price is then fed into the grid into DSM systems which adjust load usage appropriately. The bold red lines in  Fig. \ref{f:diagram} highlight the feedback relationship between price and demand. The scope of work is in the mathematical modeling of the controller and DSM system relationship. With such a model we present in the next section how attackers can exploit it.

\begin{table}[t] 
	\centering
	\begin{tabular}{ |l|l| }
		\hline
		\multicolumn{2}{|c|}{Simulation Parameters} \\
		\hline
		$ N $ & 200 homes \\
		$\epsilon_{dsm}$ & -1 \\
		$L_{target}$ & 200 kWh \\
		\hline
	\end{tabular}
	\caption{Simulation parameters used in case studies.}
	\label{t:parameters}
\end{table}

\begin{figure*}[t]
	\begin{subfigure}{.33\textwidth}
		\centering
		\includegraphics[scale=0.45]{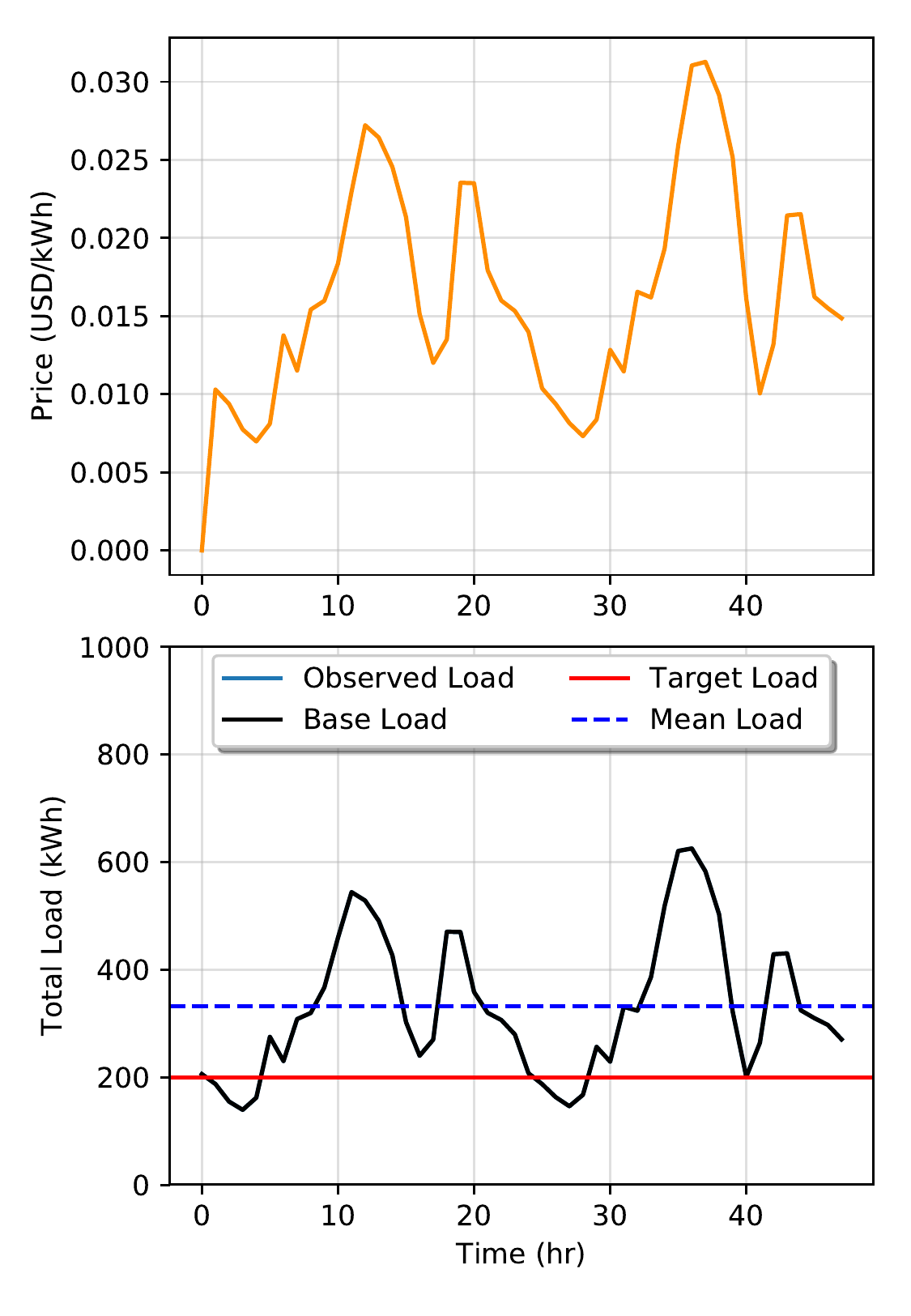}
		\caption{}
		\label{f:a1}
	\end{subfigure}
	\begin{subfigure}{.33\textwidth}
		\centering
		\includegraphics[scale=0.45]{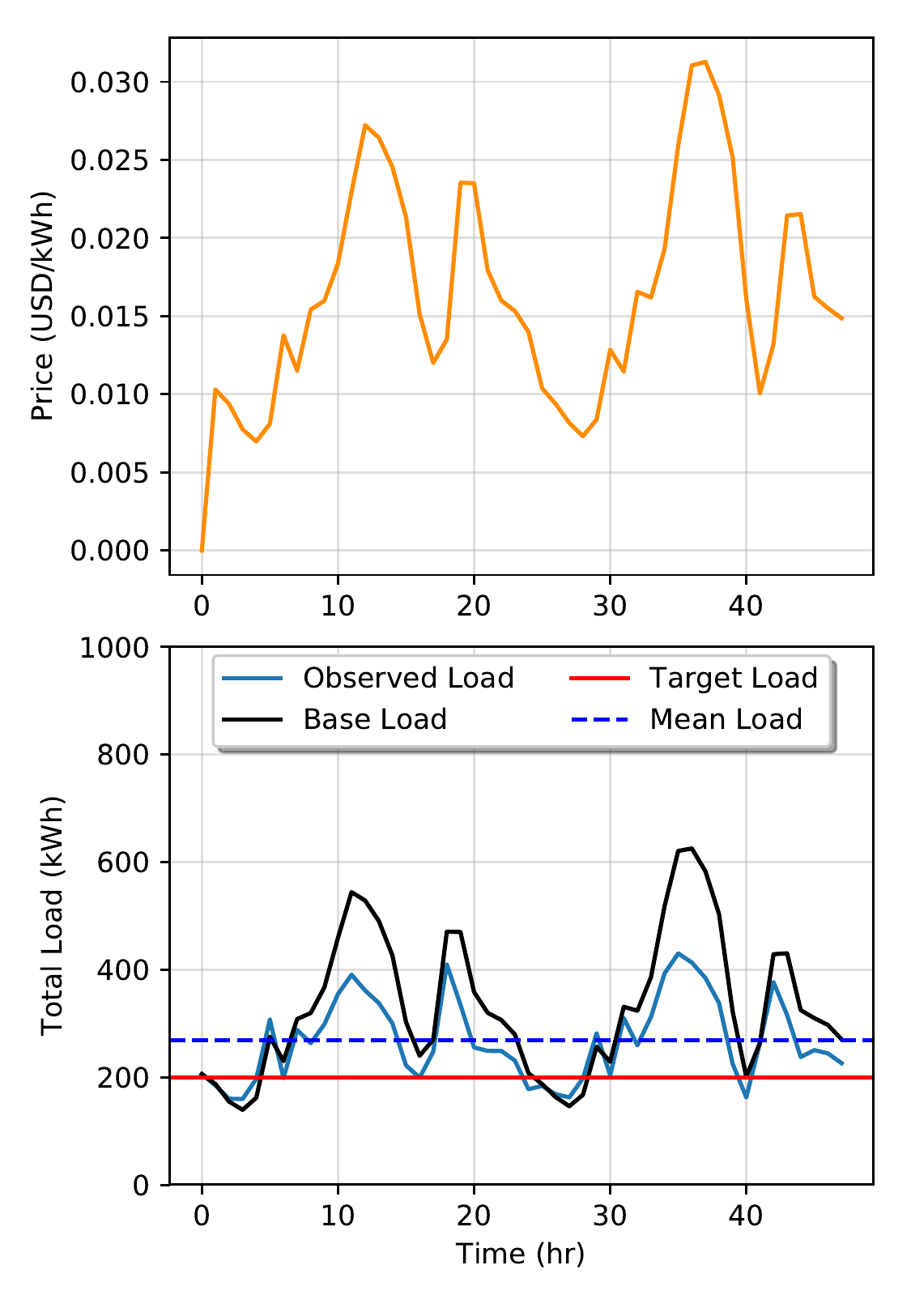}
		\caption{}
		\label{f:a2}
	\end{subfigure}
	\begin{subfigure}{.33\textwidth}
		\centering
		\includegraphics[scale=0.45]{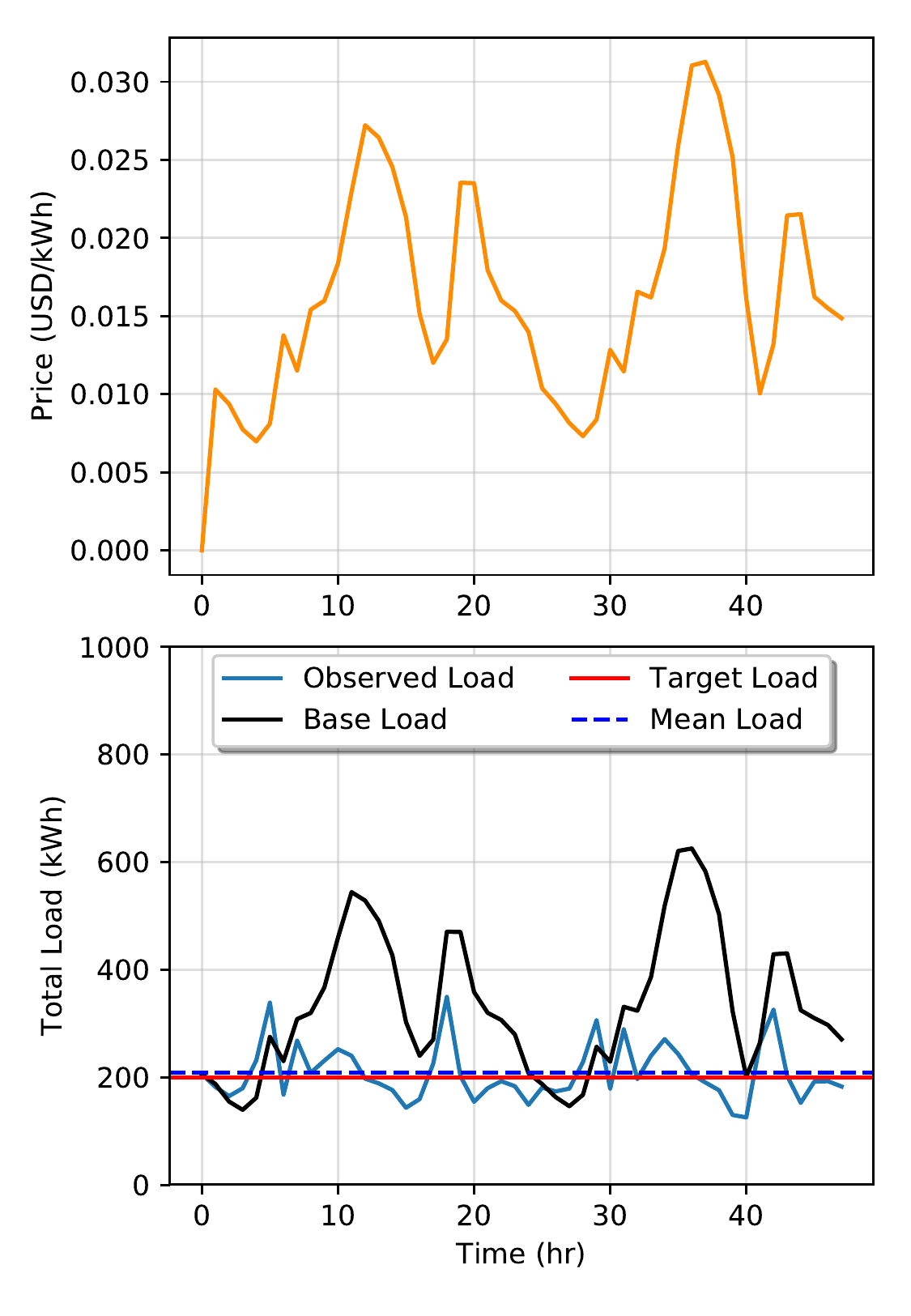}
		\caption{}
		\label{f:a3}
	\end{subfigure}
	\caption{Simulation of price-load interaction with Goal 1 with DSM when $\kappa_i =0, \forall_i$ (a), $\kappa_i =0.5, \forall_i$ (b), and $\kappa_i =0.99, \forall_i$ in (c).}
	\label{f:goal0}
\end{figure*}

\begin{figure*}[t]
	\begin{subfigure}{.33\textwidth}
		\centering
		\includegraphics[scale=0.45]{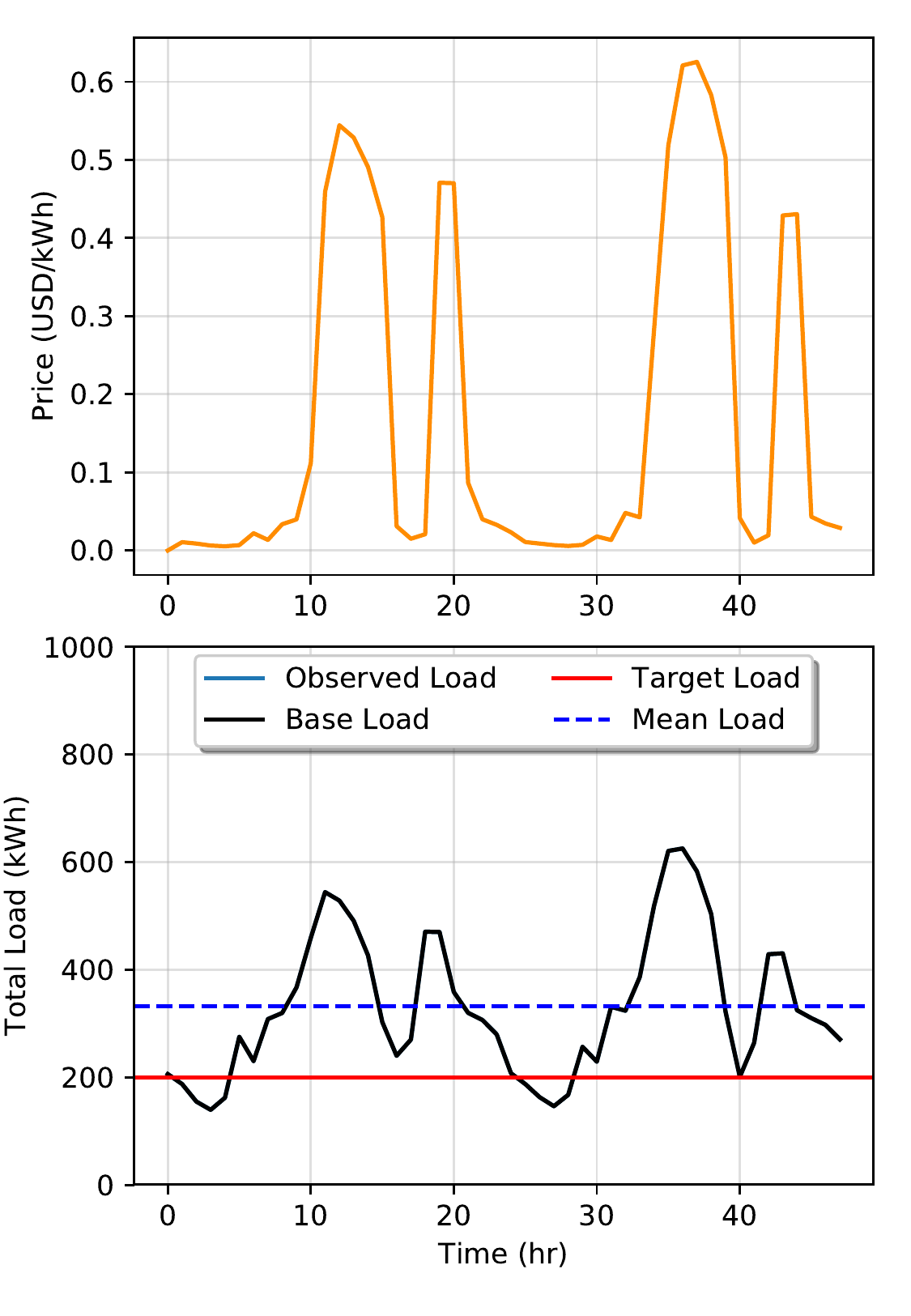}
		\caption{}
		\label{f:b1}
	\end{subfigure}
	\begin{subfigure}{.33\textwidth}
		\centering
		\includegraphics[scale=0.45]{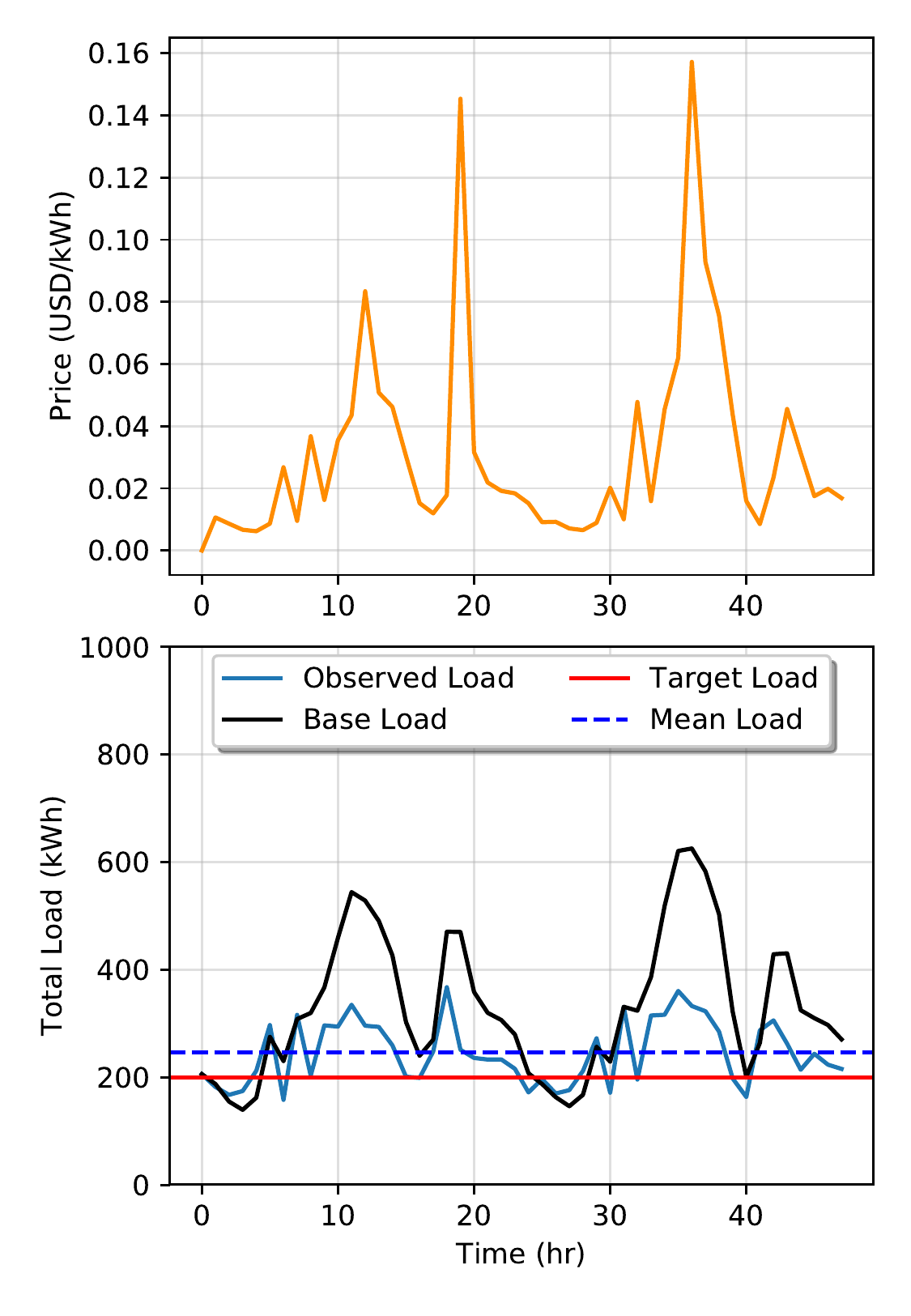}
		\caption{}
		\label{f:b2}
	\end{subfigure}
	\begin{subfigure}{.33\textwidth}
		\centering
		\includegraphics[scale=0.45]{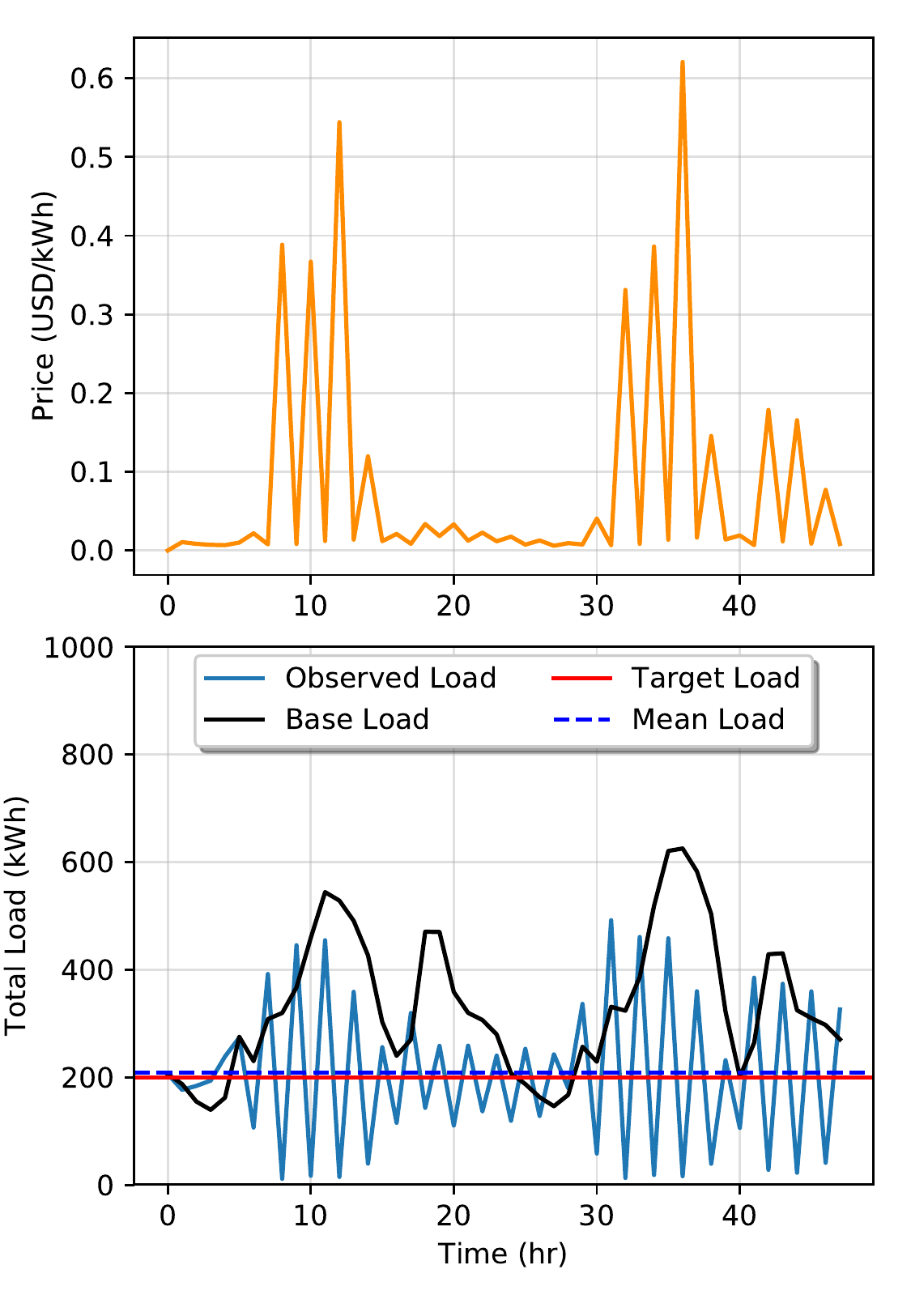}
		\caption{}
		\label{f:b3}
	\end{subfigure}
	\caption{Simulation of price-load interaction with Goal 2 with DSM when $\kappa_i =0, \forall_i$ (a), $\kappa_i =0.5, \forall_i$ (b), and $\kappa_i =0.99, \forall_i$ in (c).}
	\label{f:goal1}
\end{figure*}

\subsection{Simulation Parameters and Assumptions}

For experimental purposes, in modeling individual user loads we set $ \phi_{t,i} $ equal to simulated bootstrapped user load profiles defined in Section \ref{s:bootstrap}. We assume that prices, user preferences $ \theta_{t,i} $, and $ \epsilon^{d}_{t,i} $ have been absorbed in the calculation of the simulated load series. Thus we use $ \phi_{t,i} $ as a reference point to how much electricity a user wants to consume without the influence of a DSM program. We model the task of the ISO as modifying $ \phi_{t,i} $ to some desired load levels. Furthermore, we include the following assumptions in our modeling and simulation. We assume the ISO can define $\epsilon^{dsm}$ for each household and we set it as a constant for all customers and time. For additional simplicity, we model $\kappa_i$ also invariant in time, but it may vary per customer. Lastly, through AMI, we assume ISO the can obtain an estimated reading of $ \Phi_{t-1} $ but not of individual user $ \phi_{t-1,i} $ to preserve privacy. This way the ISO has a time series of estimated non-DSM load demand in order to provide predictions $\hat{\Phi}_t$.

For all case studies in the rest of our paper, we use the UMass Smart* dataset (2017 release) to bootstrap simulate residential load as described in Section \ref{s:Microgrid}. We simulate a micro-grid of $N = 200$ residential homes for a time period of $T$ to obtain a $phi$ distribution that defines a base load profile for each home $i$ at each time $t$. We set the DSM demand elasticity for each user to $ \epsilon^{dsm} = -1, \forall_i$ to allow the DSM component of customer load to be sensitive enough to price changes. For all case studies we set a simple flat target load of $L_{target} = 200$ kWh. We note, however, that with our pricing formulation in Eq. \ref{e:P2} we can model any DSM goal from peak load reduction to flexible load shaping. These simulation parameters are summarized in Table \ref{t:parameters}. For forecasting $\hat{\Phi}_t$ we use the naive persistence prediction method.

In Figures \ref{f:goal0} and \ref{f:goal1} we demonstrate a price-load feedback simulation for a period of $ T = 48 $ hrs under Goals 1 and 2 with parameters defined in Table \ref{t:parameters}. Under the Goal 1 scenario when $\kappa_i = 0, \forall_i$, shown in Fig. \ref{f:goal0}(a), we see that prices range from \$0.01 to \$0.03 per kWh. As expected with no DSM program to influence demand, the observed load is equal to the base load. The same is observed in Fig. \ref{f:goal1}(a) when the simulation is run with the Goal 2 RTP model. The only difference under this scenario is the price range which is exceptionally high ranging from \$0.01 to \$0.6 per kWh. This is expected in this scenario since the ISO is attempting to set prices to maximize the effect on DSM customers, which there are none, but this is unknown to the utility. With no DSM the mean observed load is 332 kWh.

Next, we rerun the simulation setting $\kappa_i = 0.5, \forall_i$. In the Goal 1 scenario in Fig. \ref{f:goal0}(b) we see that the base load was reduced with a mean observed load of 269 kWh. In Fig. \ref{f:goal1}(b) the same simulation is run with Goal 2. Here the mean observed load was further reduced to 246 kWh, but large price spikes occur at peak load times. Finally, we run simulations for $\kappa_i = 0.99, \forall_i$ to study the effects of high penetration of DSM. In simulating both goals, shown in Figures \ref{f:goal0}(c) and \ref{f:goal1}(c), the mean observed load is reduced to 208 kWh, which is very close to the target load. However, in Goal 2 we see great resonating feedback affect occur when prices spike very high. RTP increases as a response to large values in observed load. Then when load decreases to low levels prices decrease cause load to spike more during the next time step. While under Goal 1 this is not observed. The higher prices set in Goal 2 would see a large cost to DSM participating customers.

\begin{figure*}[t]
	\begin{subfigure}{.33\textwidth}
		\centering
		\includegraphics[scale=0.45]{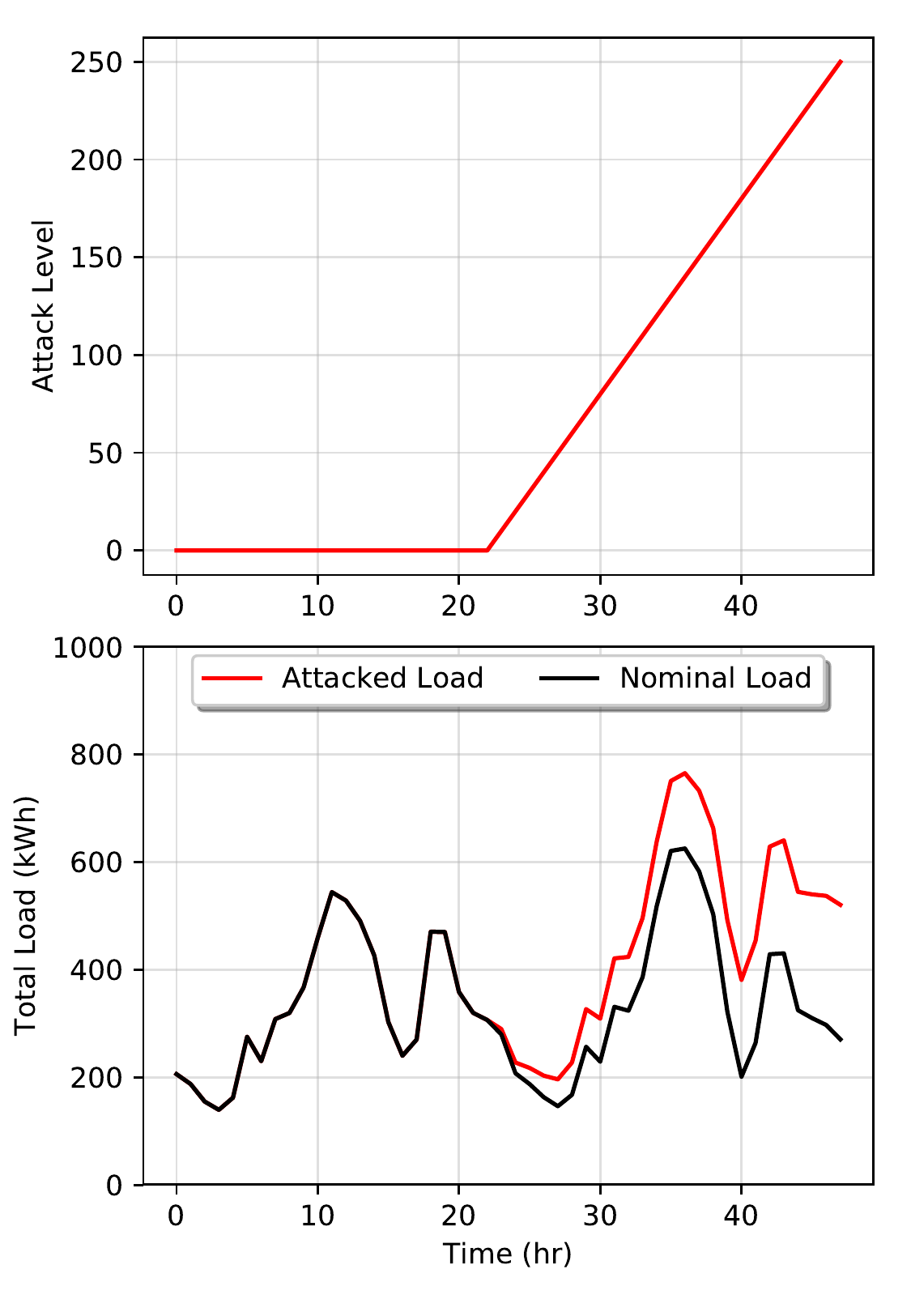}
		\caption{}
		\label{f:b1}
	\end{subfigure}%
	\begin{subfigure}{.33\textwidth}
		\centering
		\includegraphics[scale=0.45]{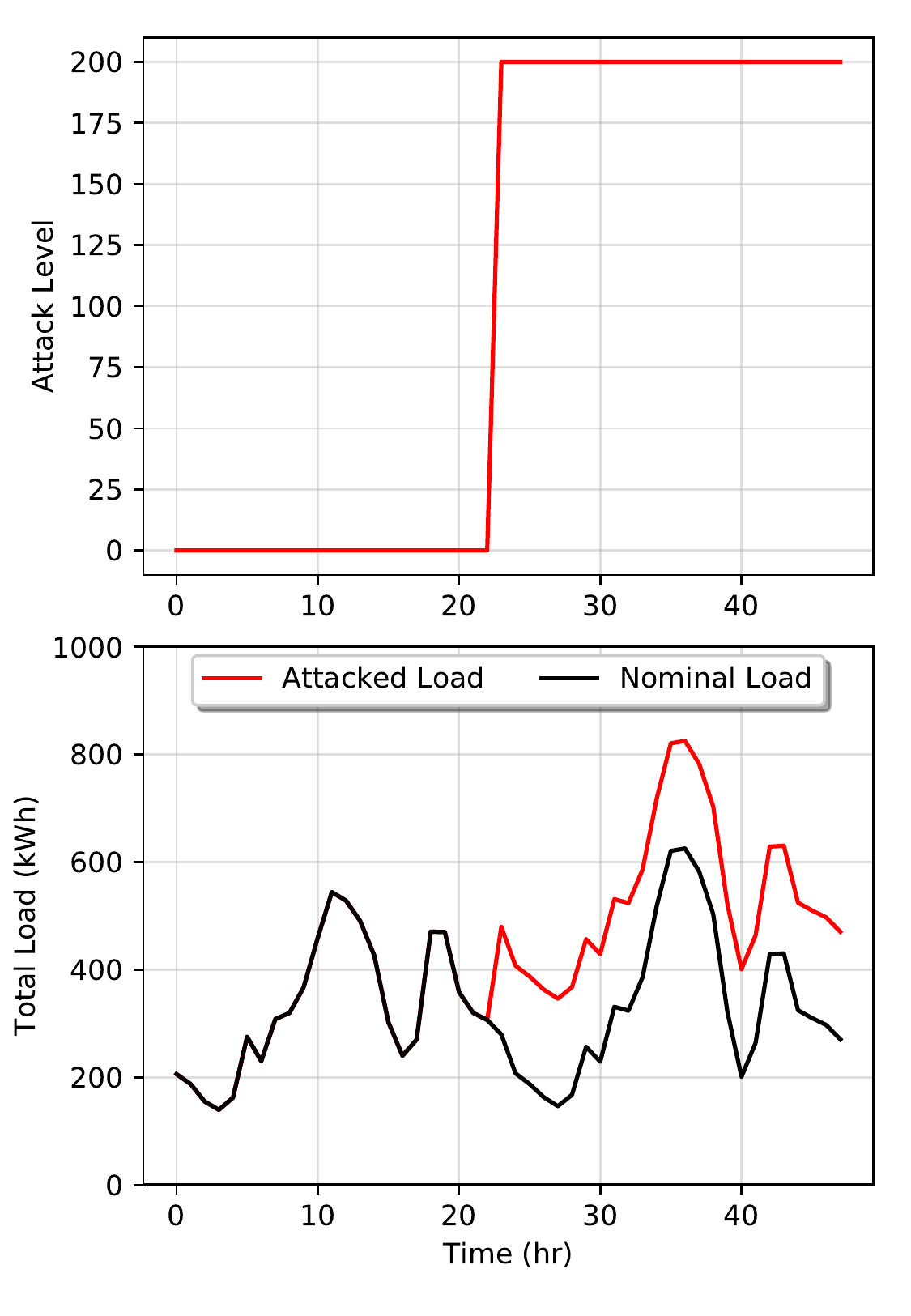}
		\caption{}
		\label{f:b2}
	\end{subfigure}
	\begin{subfigure}{.33\textwidth}
		\centering
		\includegraphics[scale=0.45]{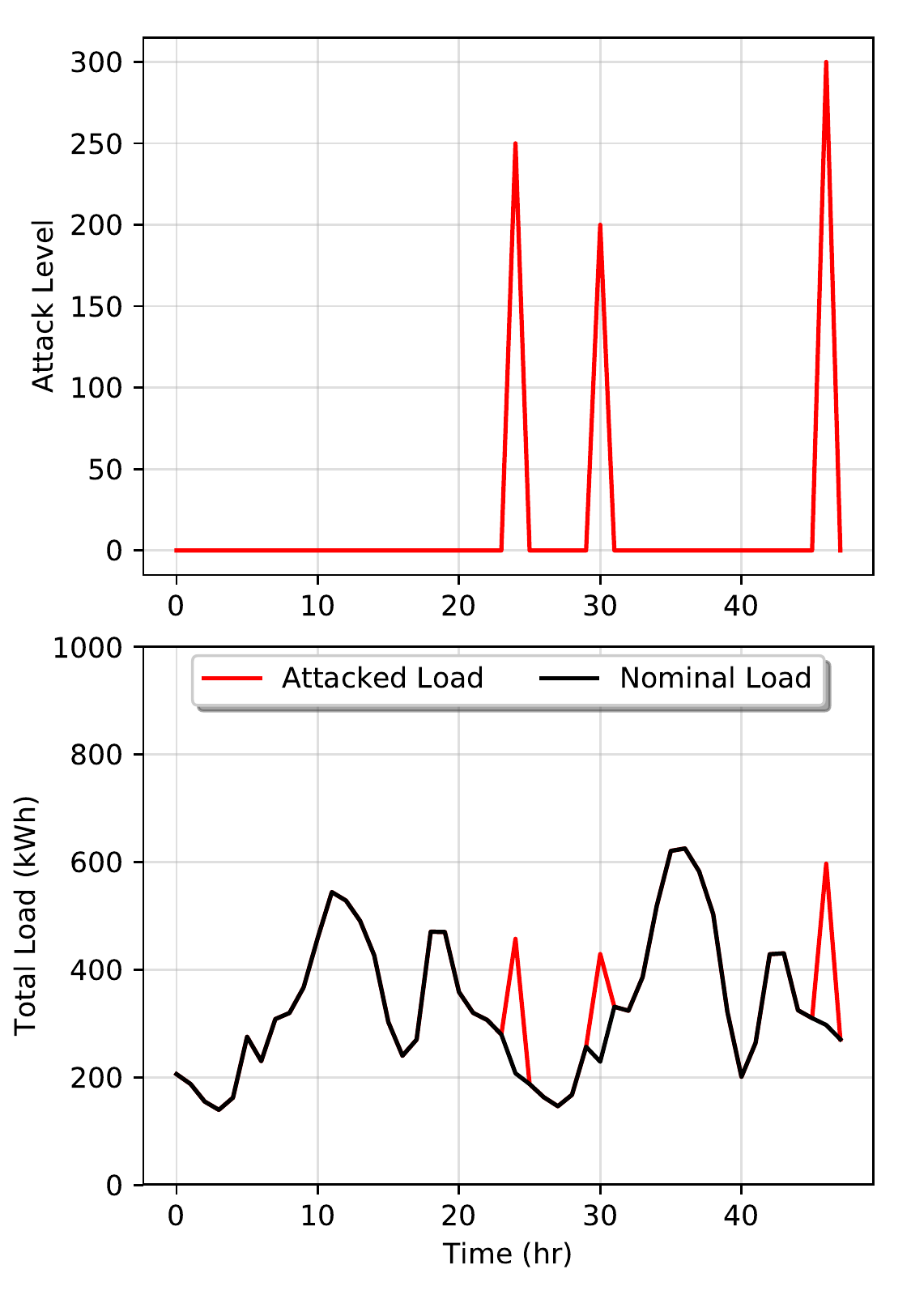}
		\caption{}
		\label{f:b3}
	\end{subfigure}
	\caption{Examples of different type of DSM attacks: (a) ramp attack, (b) sudden attack, and (c) point attack.}
	\label{f:attack_types}
\end{figure*}

\begin{figure}[t]
	\centering
	\includegraphics[scale=0.7]{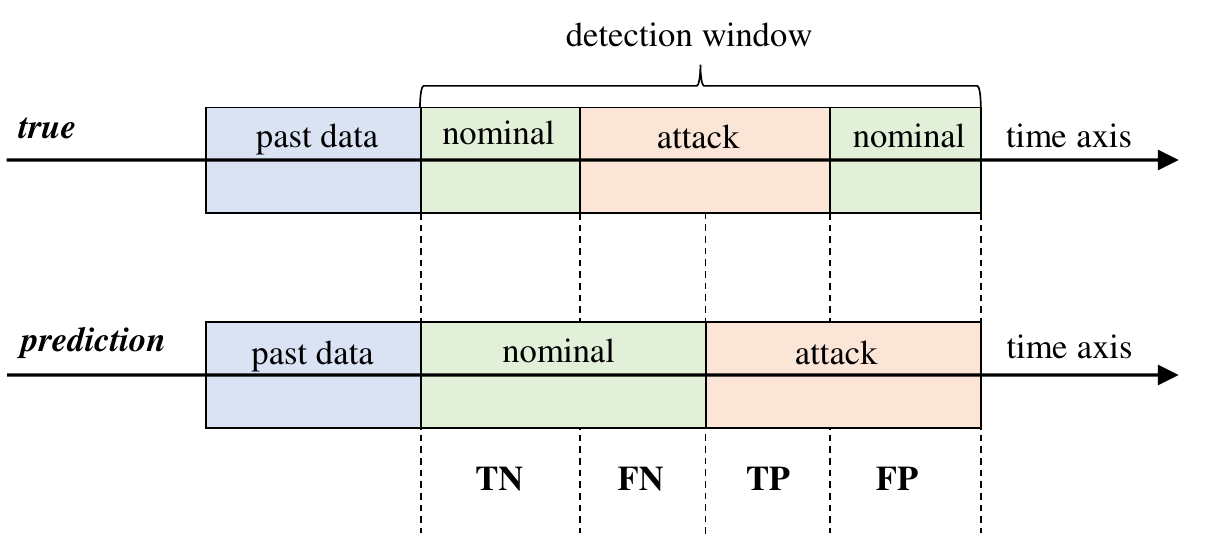}
	\caption{Confusion matrix imposed on a time axis of attack predictions vs true observations.}
	\label{f:matrix}
\end{figure}

\section{DSM Attack Models} \label{s:attack}

An attacker can exploit the feedback between the customer and utility in determining RTP and load usage by cyber DSM programs by injecting false price or corrupted load data into the feedback loop. The attack exploitations we study here are different from the false data injection attacks studied in other smart grid papers. Most false data injection attack works \cite{liang2017review} study the compromise in energy management systems to alter power state estimates by the utility operator. In our case, we study attacks that aim to alter a users load profile by exploiting cyber DSM vulnerabilities. 

For modeling attacks on a cyber DSM managed micro-grid we assume that the attacker compromises a subset of all the $ N $ customers, we denote this subset as $ \mathcal{A} $, for an attack period $ t \in T_a $. We study two modes of attacks: false pricing data injection attacks in which a compromised user receives manipulated pricing information, and a direct load manipulation attack in which the appliances of the compromised customer are under the control of the attacker. When communication encryption is broken with an AMI, then a pricing data injection attack can occur. By hacking into a cyber DSM load controller, or directly hacking into smart appliances, then a direct load manipulation can occur by altering a users load profile. The two modes of attacks are outlined below.

\begin{itemize}
	\item \textbf{False Pricing Data Injection}: The attacker can manipulate prices $P_t$ received by each compromised costumer $i \in \mathcal{A}$, and the received price $P_{t}^{i}$ can be different for various customers in order to achieve the attacker's desired effect:
	\begin{equation*} 
	P_{t,i}^{a} = P_{t,i} + a_{t,i}^{P}, \forall_i \in \mathcal{A}, t \in T_a
	\end{equation*}
	This has the affect of compromising the demand response of a customer in the following way:
	\begin{equation*}
	l_{t,i}^{a^P} = (\kappa_i \phi_{t,i}) (a_{t,i}^{P})^{\epsilon^{dsm}} + (1-\kappa_i) \phi_{t,i} 
	\end{equation*}
	\item \textbf{Direct Load Manipulation}: The attacker can manipulate the load of each compromised customer $l_{t,i}, i \in D$ directly:
	\begin{equation*} 
	l_{t,i}^{a^L} = l_{t,i} + a_{t,i}^{L}, \forall_i \in \mathcal{A}, t \in T_a 
	\end{equation*}
	Under both attack modes, we would get a compromised aggregate load which may include one or both attacks occurring simultaneously
	\begin{equation*} 
	L_{t}^{a} = \sum_{i=1}^{N} l_{t,i}^{a^P} + l_{t,i}^{a^L}.
	\end{equation*}
\end{itemize}

\noindent These two modes of attack are equivalent as they both affect a customers load response as long a part of the load is under cyber DSM control that is sensitive to price changes.

\bigskip
\noindent \textbf{Theorem 1.} Given a set of customers $ \mathcal{A} $ compromised by the attacker, there always exist a direct load manipulation attack such that all customers behave the same as a pricing data injection attack and vice versa for $\kappa_i > 0, \forall_i, t \in T_a $.

\begin{proof}
	Setting both attacks to have the same load, then the attack load is set as follows
	\begin{equation*} 
	a_{t,i}^{L} = (\kappa_i \phi_{t,i}) (a_{t,i}^{P})^{\epsilon^{dsm}} + (1-\kappa_i) \phi_{t,i},
	\end{equation*}
	and the equivalent attacked price is
	\begin{equation*} 
	a_{t,i}^{P} = \left( \frac{a_{t,i}^{L}-(1-\kappa_i) \phi_{t,i}}{\kappa_i \phi_{t,i}}\right)^{1/\epsilon^{dsm}}
	\end{equation*}
	\noindent If  $\kappa_i = 0$ then the two attack modes are not equivalent since a price attack will have no affect on customer load.
\end{proof}

\noindent Since false pricing data injection and direct load manipulation attacks are equivalent, we focus only on direct load manipulation attack analysis. 

There are different goals an attacker can have to harm the power grid or exploit it. For example, an attacker can cause chaotic metering by messing the metering data transmission, efficiency loss of the energy provided by causing greater load volatility, or the energy system failure by overloading the power lines or devices. The focus of this work is efficiency loss by increasing user loads through direct load manipulation. In this scenario, we introduce three possible types of load attacks. A ramp attack, sudden attack, and point attack. These type of attacks are shown in Fig. \ref{f:attack_types} wherein plot (a) an attacker gradually increases a users load over time. In plot (b) an attacker suddenly ramps up the power usage to a specified level, and in plot (c) we demonstrate a point attack where the attacker increases loads only for specific hours.

\begin{figure}[t]
	\centering
	\includegraphics[scale=0.55]{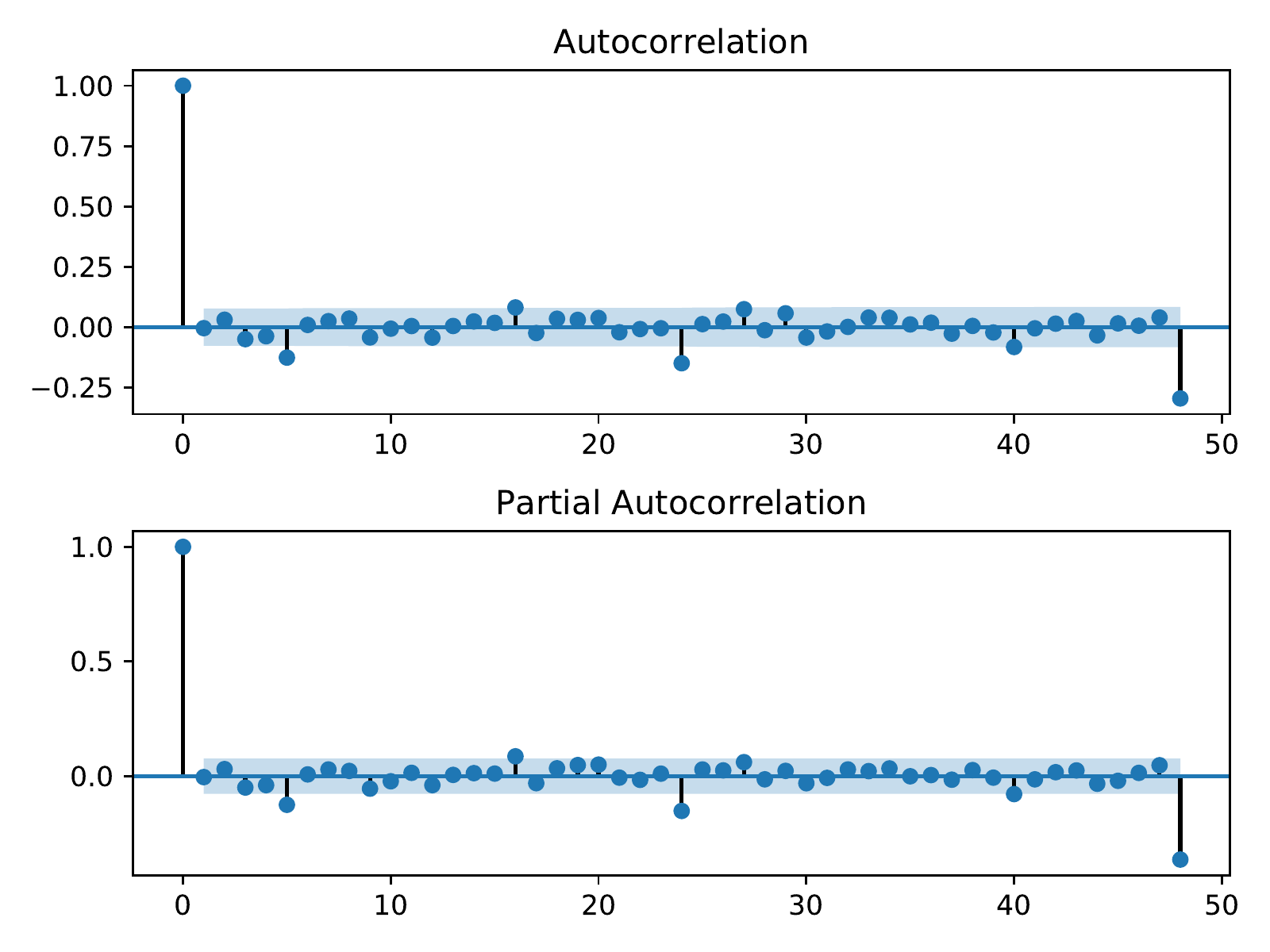}
	\caption{ACF and PACF fplots of the residual series between the SARIMA fit tp training data. From the plots we observe that the residuals are stationary.}
	\label{f:acf_residuals}
\end{figure}

\begin{figure}[t]
	\centering
	\includegraphics[scale=0.55]{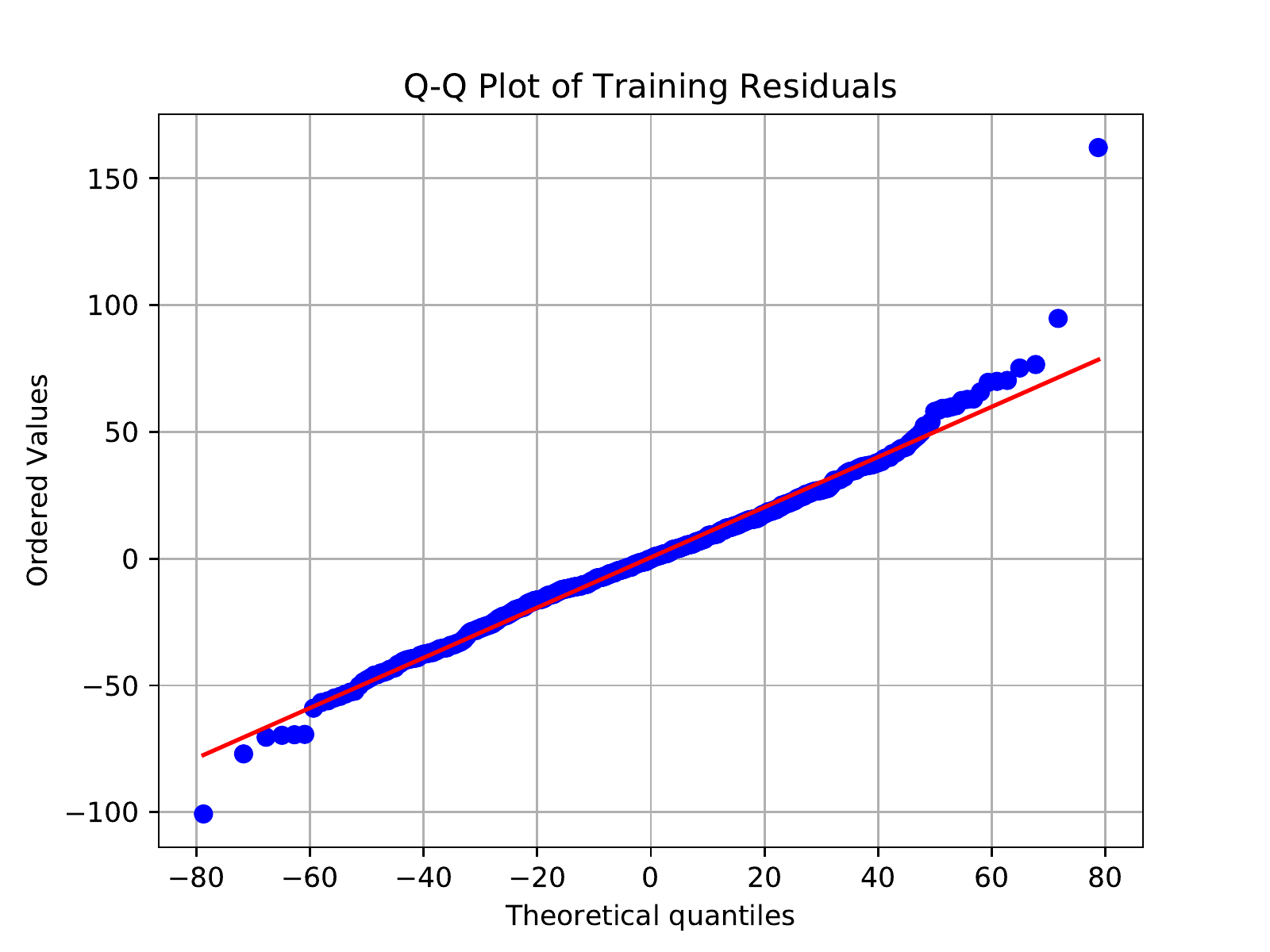}
	\caption{Q-Q plot of the residual series between the SARIMA fit tp training data. From the plot we observe that for extreme quantiles the distribution is not Gaussian.}
	\label{f:qq_residuals}
\end{figure}

\section{Attack Detection}

Here we outline sequential and supervised learning-based methods for attack detection. The sequential detections methods are the one-sided cumulative sum (CUSUM) test \cite{tartakovsky2014sequential} and windowed generalized likelihood ratio test (GLRT) \cite{key1993fundamentals}. Both these methods take as an input a residual time series that is the output of applying a SARIMA filter to load observations. If enough past observations of load data that is labeled as nominal or under attack are collected, then detection of attacks can be made by training supervised learning classifiers. Supervised learning algorithms have been broadly adopted to the smart grid literature for monitoring and detecting cyber attacks on power systems \cite{hink2014machine, ozay2016machine, esmalifalak2017detecting}. Here we employ several supervised learning methods for detecting attacks on cyber-DSM systems. It is unlikely that an ISO can collect high-quality attack data due to the lack of such attacks occurring. However, such data can be simulated. Using past load data we simulate direct load manipulation attacks by creating different types of attacks as shown in Fig. \ref{f:attack_types}.

\subsection{Sequential Detection Methods}

In sequential change point detection, a series does not have a fixed length. Instead, observations are received and processed sequentially over time. When an observation has been received, a decision is made about whether a change has occurred in the state based only on the observations which have been received so far or within a fixed past window size. If no change is detected, then the next observation in the sequence is processed. The sequential formulation allows sequences containing multiple change points to be easily handled. Sequential change point detection can be applied in the case of attack detection to identify if a load time series has been compromised. If an attack is flagged, then an ISO can take appropriate actions to prevent further damage to the grid.  

Under the Sequential Detection paradigm we collect observations an apply a whitening filter to produce a residual series with the assumption that it is white Gaussian noise. If an additive attack $ A_t > 0 $ is present for observation $t$, this will cause a definite shift in the mean of the residual series. This detection problem can thus be stated as deciding if a null hypothesis $ \mathcal{H}_0 $ is true, where the residual series has zero mean and known variance (invariant in time and estimated from a sample population), or if the alternative hypothesis $ \mathcal{H}_1 $ is true which states that the examined series has some mean not equal to zero thus being under attack. This can be modeled as a hypothesis test, and for the GLRT and CUSUM detectors this translates to 
\begin{equation*}
\begin{split}
\mathcal{H}_0 & : x_t \myeq \mathcal{N}(0,\sigma^2), t = 1,...,N \\
\mathcal{H}_1 & : x_t \myeq \mathcal{N}(A_t,\sigma^2), A_t>0, t = 1,...,N
\end{split}
\end{equation*}

\noindent For the GLRT detector, to simplify implementation, we model the attack as if it were constant $A>0$ but unknown. To produce a residual series a SARIMA multi-step forecast for $ t = 1,...,N $ is made before the detection period. This forecast is conducted using a past window of training data that was not under attack. The forecast is made for time $ t $ to $ t+k $. These predictions are then subtracted from the incoming observations to produce a residual time series which is then fed as input into the GLRT and CUSUM detectors. 

The generalized likelihood ratio with a threshold to decide $ \mathcal{H}_1 $ is defined as
\begin{equation*}
\frac{p(\textbf{x};A,\mathcal{H}_1)}{p(\textbf{x};\mathcal{H}_0)} = \frac{\frac{1}{\sqrt{2\pi \sigma^2}} \exp^{-\frac{1}{2\sigma^2}\sum_{t=1}^{N}(x_t - A)^2}}{\frac{1}{\sqrt{2\pi \sigma^2}} \exp^{-\frac{1}{2\sigma^2}\sum_{t=1}^{N}x_{t}^{2}}} \underset{H_0}{\overset{H_1}{\gtrless}} \gamma
\end{equation*}

\noindent Taking the log of both sides simplifies the results to
\begin{equation*}
T(\textbf{x}) = \frac{1}{N}\sum_{t=1}^{N} x_{t} \underset{H_0}{\overset{H_1}{\gtrless}} \frac{\sigma^2}{NA} \ln \gamma + \frac{A}{2} = \gamma'
\end{equation*}

\noindent The threshold is then found by
\begin{equation*}
\begin{split}
P_{FA} = P(T(\textbf{x})>\gamma';\mathcal{H}_0) = Q\left( \frac{\gamma'}{\sqrt{\sigma^2 / N}}\right) \\
\Rightarrow \gamma' = \sqrt{\frac{\sigma^2}{N}} Q^{-1}(P_{FA})
\end{split}
\end{equation*}

\noindent where $N$ is the size of our window and $\sigma$ is estimated from past training data used to produce the SARIMA forecasts.

%

A CUSUM test is a control chart, that is used to monitor the mean of a process based on samples taken from past data at specific time intervals. It is a class of non-linear stopping rules for structural changes. Given information of current and previous samples, a CUSUM test relies on the specification of a target value $h$ and a known or reliable estimate of the standard deviation $\sigma$ the process. The CUSUM test typically signals an out of control or anomalous process by an upward or downward drift of the cumulative sum until it crosses the target threshold. For attack detection, if the mean of the load series shifts above the target threshold, we then assume the grid is under attack. 

We define the CUSUM detector as follows. Taking the residual series $ x_t = y_t - E_{t-1}[y_t]$, again defined by a SARIMA forecast $ E_{t-1}[y_t]$, we define a one-sided CUSUM detector as
\begin{equation*}
g_t = \max (0, g_{t-1}+x_t - k)
\end{equation*}

\noindent where $k$ is called the reference value (sometimes also called drift) set priori to values such as 0, 0.5, or $ A/2 $ if the size of $ A $ is known in advance. When $g_t = 0$ then we define the change time as $t_c = t$, and when $g_t > h > 0$ we reset $g_t = 0$ and flag an alarm at time $t_a = T$. The alarm threshold is also set priori to some value based on the sample population standard deviation such as $h = 2\sigma$ where $\sigma$ is estimated from past training data used to produce the SARIMA forecasts.

 \begin{figure*}[t]
	\begin{subfigure}{.33\textwidth}
		\centering
		\includegraphics[scale=0.4]{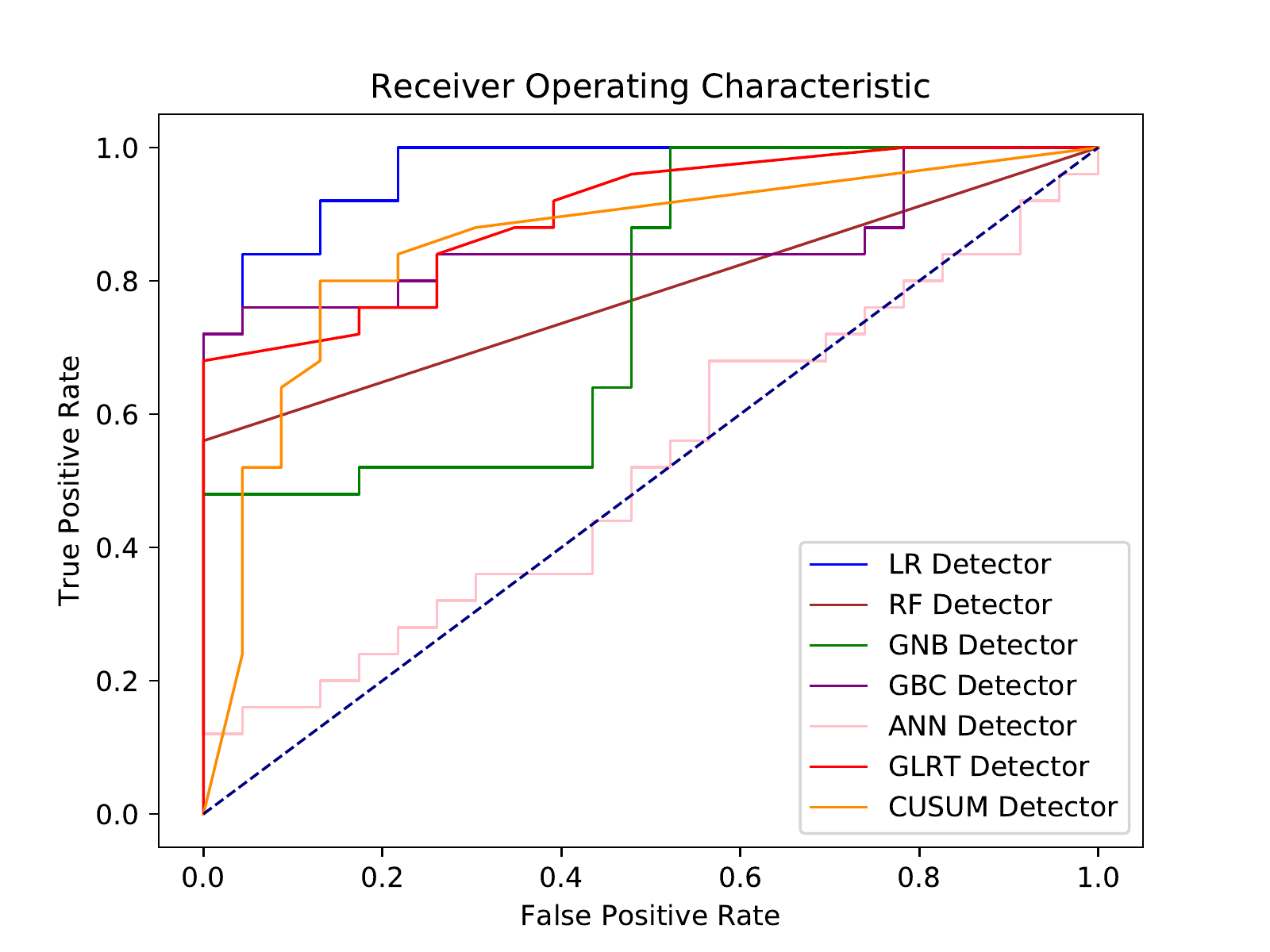}
		\caption{Type 1 attack; $\kappa=0.1$.}
		\label{f:ROC_A}
	\end{subfigure}
	\begin{subfigure}{.33\textwidth}
		\centering
		\includegraphics[scale=0.4]{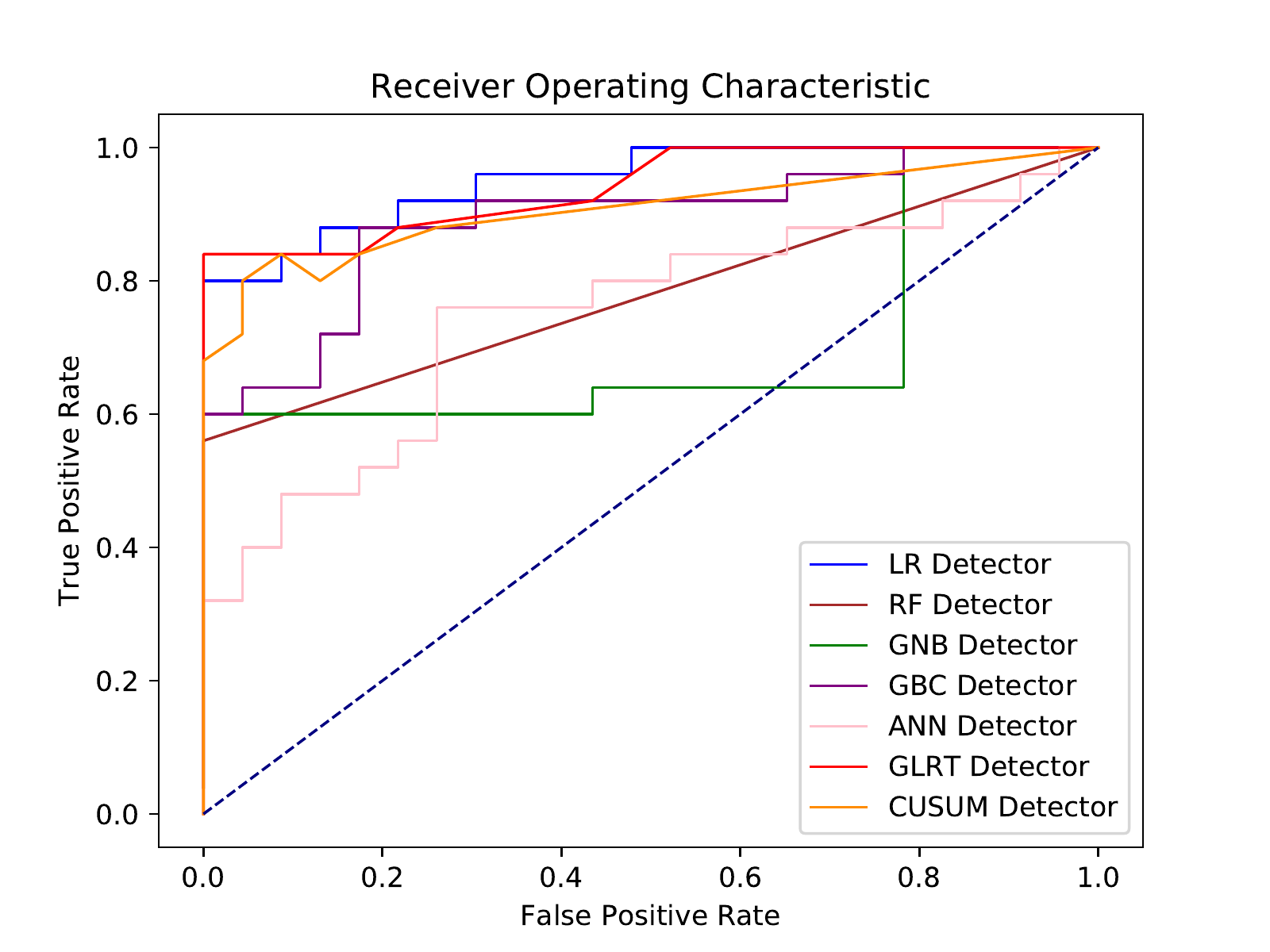}
		\caption{Type 2 attack; $\kappa=0.1$.}
		\label{f:ROC_B}
	\end{subfigure}
	\begin{subfigure}{.33\textwidth}
		\centering
		\includegraphics[scale=0.4]{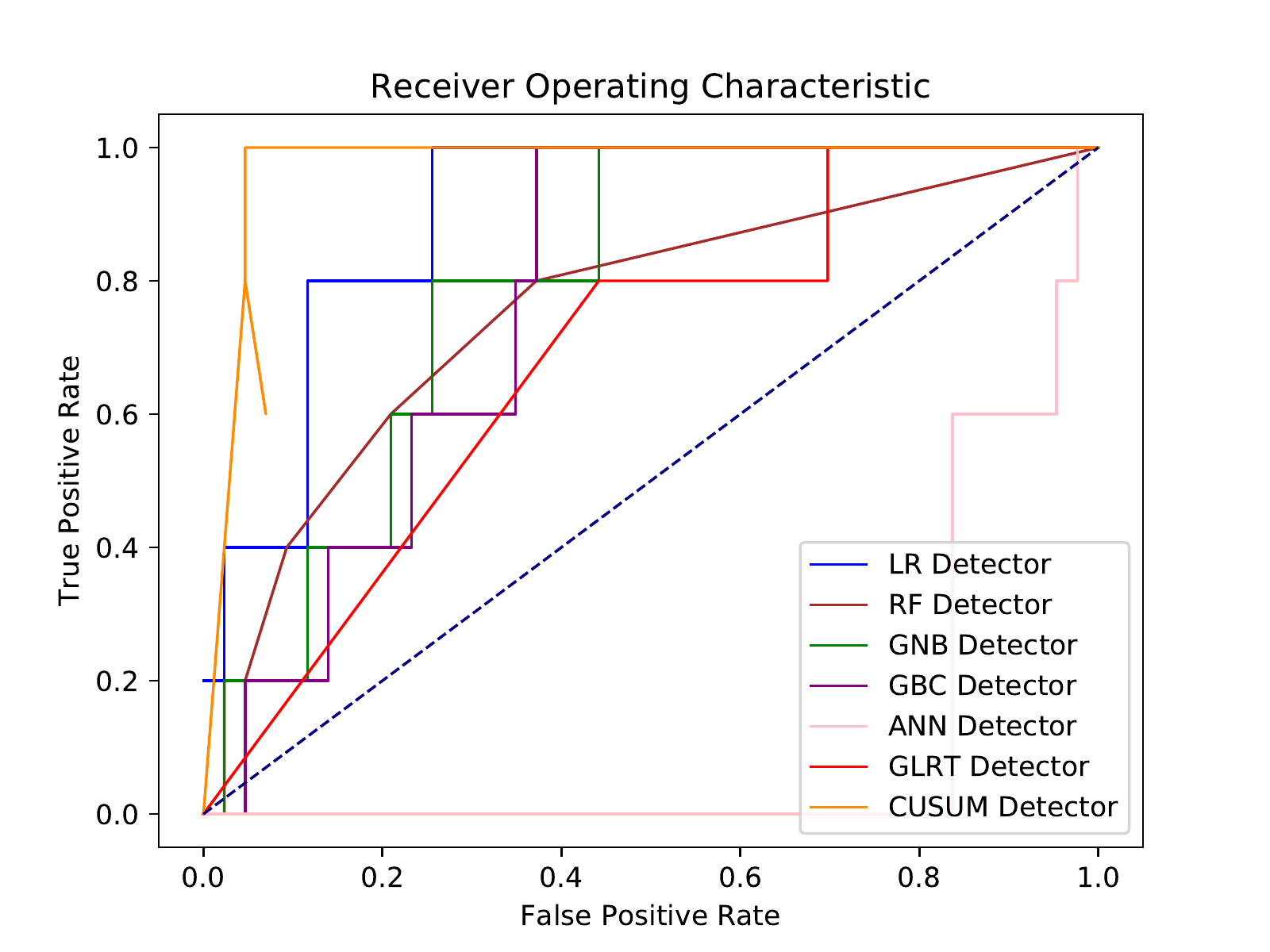}
		\caption{Type 3 attack; $\kappa=0.1$.}
		\label{f:ROC_E}
	\end{subfigure}
	
	\begin{subfigure}{.33\textwidth}
		\centering
		\includegraphics[scale=0.4]{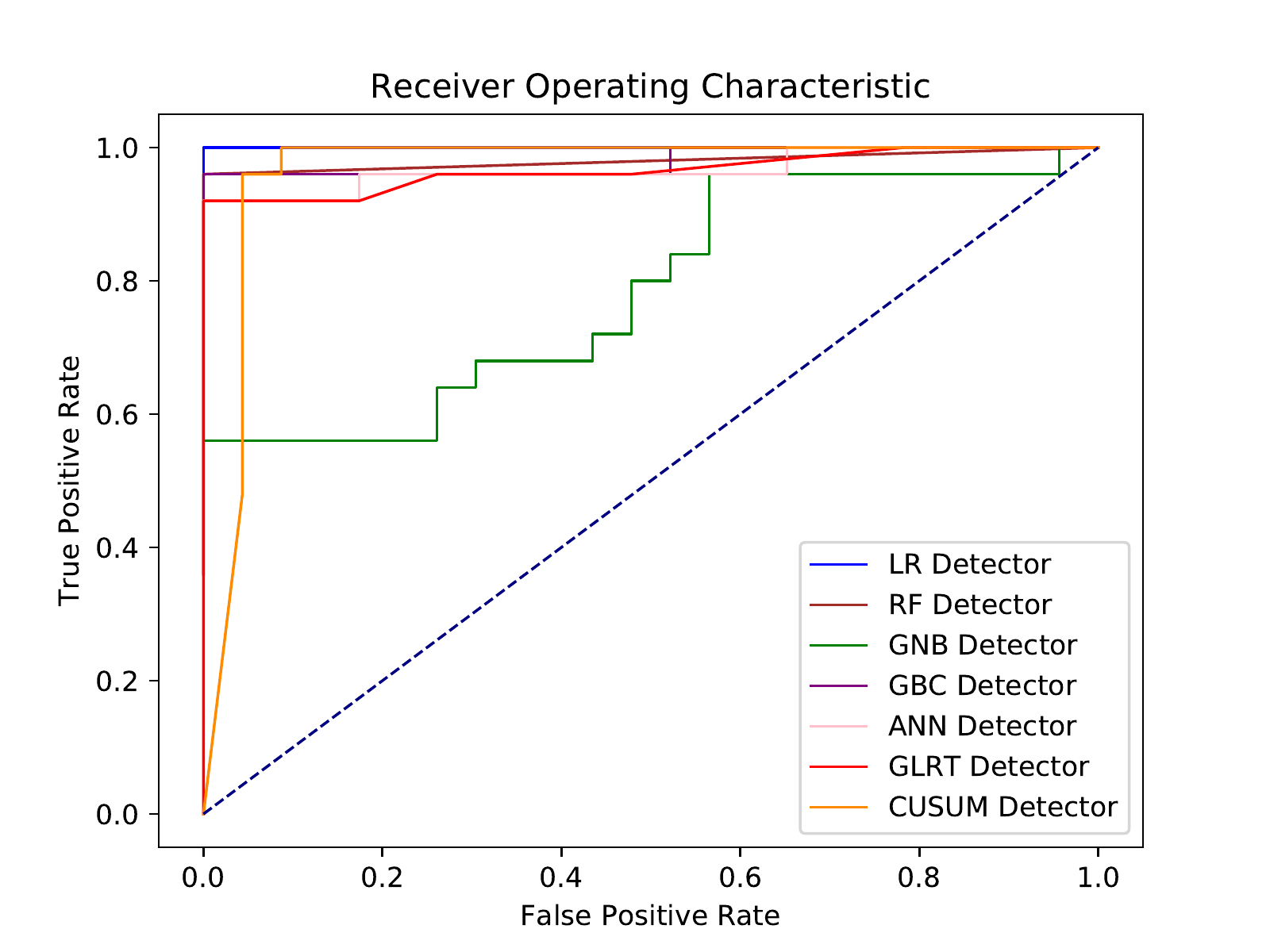}
		\caption{Type 1 attack; $\kappa=0.9$.}
		\label{f:ROC_B}
	\end{subfigure}
	\begin{subfigure}{.33\textwidth}
		\centering
		\includegraphics[scale=0.4]{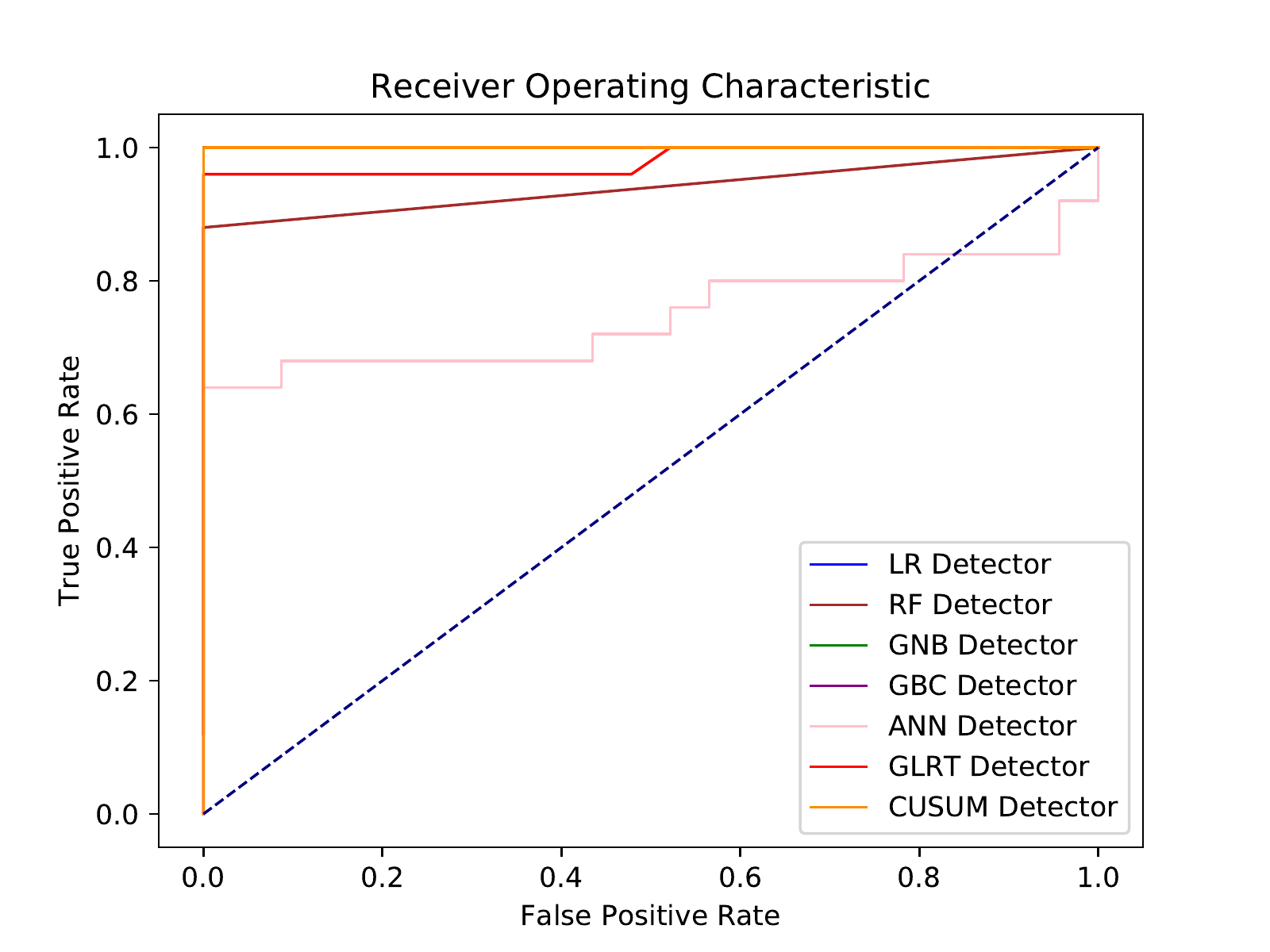}
		\caption{Type 2 attack; $\kappa=0.9$.}
		\label{f:ROC_D}
	\end{subfigure}
	\begin{subfigure}{.33\textwidth}
		\centering
		\includegraphics[scale=0.4]{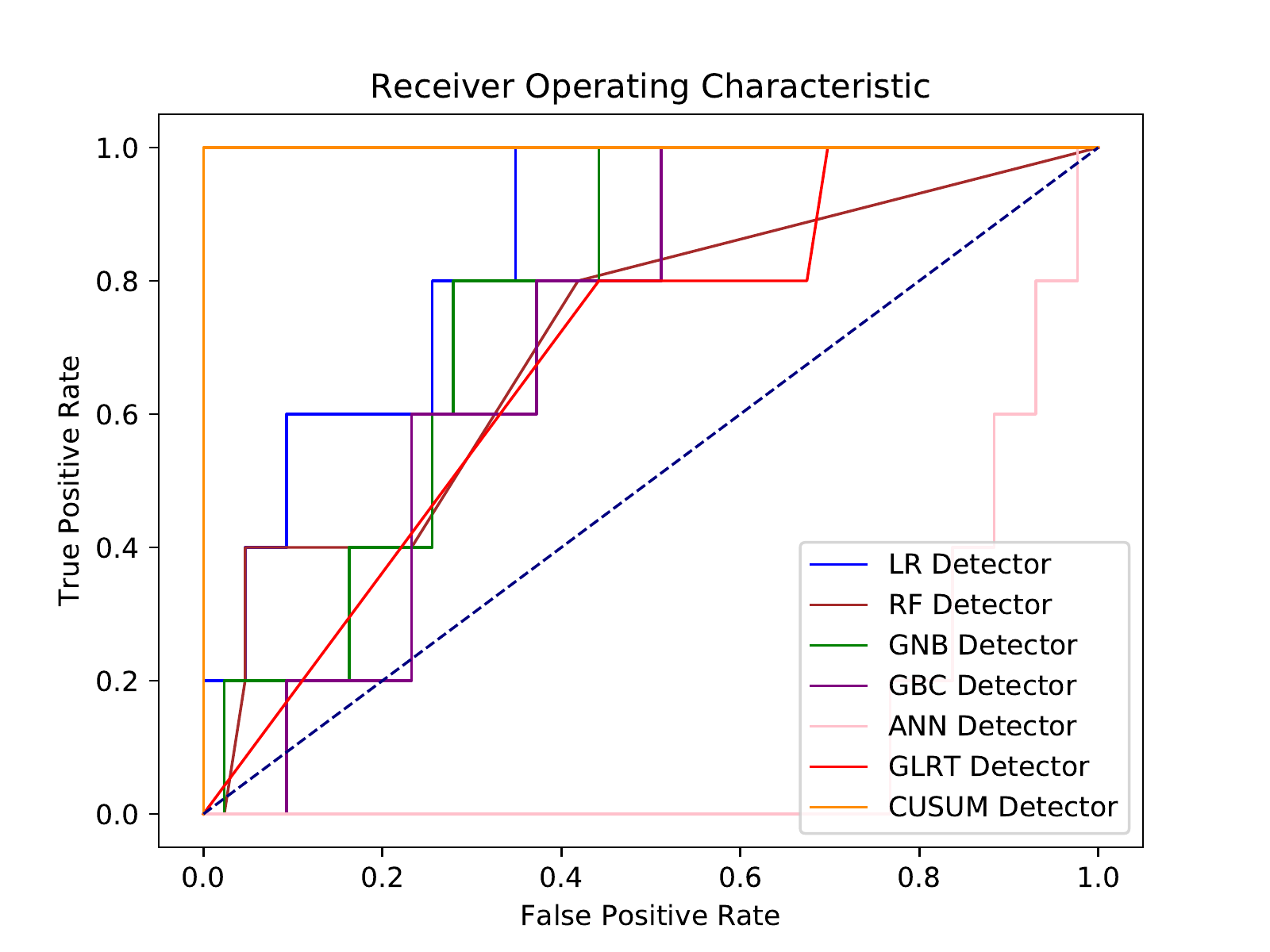}
		\caption{Type 3 attack; $\kappa=0.9$.}
		\label{f:ROC_F}
	\end{subfigure}
	
	\caption{ROC curves for the six different experiment scenarios.}
	\label{f:roc_curves}
\end{figure*}

\subsection{Supervised Learning Methods}

Changepoint detection could alternatively be treated as a supervised learning binary classification problem. Under this scheme, all of the change point sequences, or in our case attacks, represent one class, and all of the nominal sequences represents a second class. Supervised learning methods are machine learning algorithms that learn a mapping from input data to a target class label. Given a set of samples $\mathcal{X}=\{\mathbf{x}_i\}^{N}_{i=1}$ and a set of labels $\mathcal{Y}=\{y_i\}^{N}_{i=1}$, then the supervised learning detection problem is defined as a hypothesis function that captures the relationship between samples and labels $ f: \mathcal{X}\longrightarrow \mathcal{Y}$. A sliding window moves through the data, considering each difference between two data points as a possible change point.

An advantage of treating attack detection as a supervised learning approach is a more straightforward training phase. However, a sufficient amount and diversity of training data need to be provided to represent all of the classes. To ensure enough training data, and to prevent class label imbalance, we simulate all attack data to train our algorithms. Machine learning methods have successfully been applied several times in data injection attacks in power systems \cite{ozay2016machine, esmalifalak2017detecting}, so we analyze here their ability in detecting data attacks on cyber DSM systems. The binary classification problem for attack detection can be defined as
\begin{equation*}
y_i = \left\lbrace 
\begin{split}
1, \text{ if } A_i > 0\\
0, \text{ if } A_i = 0
\end{split}
\right. 
\end{equation*}

\noindent where $y_i = 1$ if the $i$-th observation is under attack, and  $y_i = 0$ if there is no attack. A variety of classifiers can be used for this learning problem. For detection of DSM attacks, we examine logistic regression (LR), random forests (RF), Gaussian Naive Bayes (GNB), gradient boosting classifier (GBC), and artificial neural network (ANN). We chose these classifiers because they are all very powerful and have widespread use in both industry and academia; model descriptions of these methods can be found in \cite{bishop2006pattern,james2013introduction}.

\subsection{Performance Analysis}

For the security of cyber DSM systems, the major concern is not just the detection of attacks, but also the detection of nominal data with high reliability. That is, we want a detection system that can predict not only with high accuracy but also with high precision and recall to avoid false alarms. Therefore, we measure the true positives (TP), the true negatives (TN), the false positives (FP), and the false negatives (FN). Definitions of these measures are visually shown in Fig. \ref{f:matrix}. We use these measures to calculate several main performance indicators of accuracy, precision, and recall.

We calculate accuracy as the ratio of correctly classified data points to total data points
\begin{equation*}
Accuracy = \frac{TP+TN}{TP+TN+FP+FN}
\end{equation*}

\noindent This measure provides the total classification success of the models. But alone, accuracy is not enough to get a full picture of performance. Precision is calculated as the ratio of true positive data points (attacks) to total points classified as attacks
\begin{equation*}
Precision = \frac{TP}{TP+FP}
\end{equation*}

\noindent On the other hand, recall, also known as the true positive rate (TPR), refers to the portion of attacks that were recognized correctly
\begin{equation*}
Recall = \frac{TP}{TP+FN}
\end{equation*}

\noindent Precision values give information about the prediction performance of the algorithms, whereas recall values measure  the degree of attack retrieval. For instance, a recall value equal to 1 signifies that none of the attacked measurements were misclassified as nominal.

We use one more final measure of total performance of our detectors, the receiver operating characteristics (ROC) curve. The ROC curve is an assessment that enables visual analysis of the trade-off between TPR and false positive rate (FPR). This can also be seen as the trade-off between the probability of detection and the probability of false alarm. FPR is defined as follows
\begin{equation*}
FPR = \frac{FP}{FP+TN}.
\end{equation*}

\noindent The ROC curve is constructed by plotting a two-dimensional graph with FPR on the x-axis and TPR on the y-axis at various threshold settings. A detection algorithm produces a (TPR, FPR) pair that corresponds to a single point in the ROC space. The best possible detection method would produce a point in the upper left corner, coordinate (0,1) of the ROC space. A random prediction would give a point along a diagonal line from (0,0) to (1,1). Points above this line are considered to have performance, while points below are considered with performance worse than guessing.
 
\section{Detection Experiments}

For attack simulation and detection experiments we use all the same parameters from Table \ref{t:parameters} but we varied the levels of $\kappa$ and attacks $a_t$. We simulate each of the three types of attacks as visualized in Fig. \ref{f:attack_types}, each in the form of direct load manipulation attack, under DSM participation levels $\kappa= 0.1$ and $ 0.9 $. This creates a total of six experiment scenarios. For each scenario, we simulate 28 days of training data (672 observations) and 2 proceeding days of test data (48 observations), both at a resolution of 1 hr. All training data is created nominally with no attacks. In the test data sets, the first 24 hrs of the test data are not under attack. The last 24 hours of the test data we add one of the three type of attacks. For ramp attacks, for each time step we add an attack $a_t = a_{t-1} + 5$ where the first 24 hours of the test set $a_{t=0...23} = 0$. The sudden attacks are similar except the attack is an additive constant $a_t = 150$. The last type of attack, point attacks, are $ a_{t=24} = 250, a_{t=29} = 200, a_{t=34} = 300, a_{t=37} = 100, a_{t=46} = 150 $, and $a_t =0$ everywhere else. 

For the sequential detectors, we use multi-step SARIMA forecasts to predict the next 48 hrs, and then use those forecasts to filter incoming test data for detection. Training was conducted on the training time series of 28 days of nominal data. SARIMA hyperparameters were chosen by examining lag one differenced autocorrelation function (ACF) and partial autocorrelation function (PACF) plots of the training data. These forecasts where then subtracted from the incoming test data to obtain a residual series that is input to the sequential detectors. The assumption for both detectors is that the residual series is Gaussian white noise, where the series has zero mean, and each observation is independent identically distributed from a Gaussian distribution. 

To ensure the residuals are white noise we apply the augmented Dickey-Fuller (ADF) test to check for stationarity, we examine the ACF/PACF plots to check for independence and run a Jarque-Bera test, and we examine a Q-Q plot to check if the residual series has a Gaussian probability distribution. We conduct these checks on all scenario training datasets, where we first train a SARIMA fit on them and then subtract that fit to produce the residual series. In all the training data sets, the residuals were proven to be stationary from the ADF test and independent from ACF/PACF plots. However, the Jarque-Bera test and Q-Q plots showed that the observations did not come from a Gaussian distribution. Example, ACF/PACF and Q-Q plots for $\kappa = 0.1$ are shown in Fig. \ref{f:acf_residuals} and Fig. \ref{f:qq_residuals}. Despite the residuals not being Gaussian, we still run the sequential detectors and examine their performance.

For training the supervised learning methods, for each of the six scenarios, we simulate more data to ensure proper class learning. We keep the test sets the same, but we extend each training set doubling its size. We keep the original training set, labeling it as nominal, and then make a copy of it. In the copy, we split it into three parts, each part we add one of the types of attacks at random levels. We label each observation of this set as under attack. We add the nominal and attacked training sets together to form a new training set with a total of 1,344 observations. With our training and test time series, we then create a set $\mathcal{X}$ features and $\mathcal{Y}$ labels that could be fed into the supervised learning classifiers. Each training sample $x_i \in \mathcal{X}$ is composed of 24 hours of lagged data, each hour is one feature, and each label $y_i \in \mathcal{Y}$ corresponds to a class 0 (not under attack) and class 1 (under attack). Together we get $N = 1,298$ pairs of $ (x_i,y_i) $ training samples.

We run our experiments on a computer with an Intel i7 6700 2.6 GHz, and 16 GB of RAM. Implementation of the simulations, experiments, and sequential detectors were done in Python 3.6. Implementation for SARIMA forecasts was done using the Python package Statsmodels \cite{seabold2010statsmodels}, the supervised learning methods and confusion evaluation metrics were implemented using the Scikit-Learn Python package \cite{pedregosa2011scikit}, and ROC curves were created using the Matplotlib Python package \cite{hunter2007matplotlib}. All classifiers used default hyperparameters from Scikit-Learn. We compiled all code and data used in our work, into a Python package titled LehighDSM which is publicly available on GitHub \cite{hatalis2019}.

Experimental results from the six scenarios are reported in Table \ref{t:metrics}, in the form of accuracy, recall, and precision metrics, and Fig. \ref{f:roc_curves} which showcases six ROC curves, one for each scenario. All performance measures in Table \ref{t:metrics}, were calculated using the best thresholds found in the ROC curves by searching for the shortest distance from each curve to the corner (0,1). As seen in Fig. \ref{f:roc_curves}(c,f) type 3 attacks, under any DSM level, had the lowest detection rates across thresholds with ANN being the worst performer. In Fig. \ref{f:roc_curves}(a,d) LR yielded the best performance with GNB, and again ANN being the worst. Type 2 attacks in Fig. \ref{f:roc_curves}(b,e) resulted in the best detection with CUSUM, LR, GNB and GBC having perfect performance. 

With the results from Table \ref{t:metrics}, we have the following conclusions; demonstrated findings imply that detection of attacks has a higher accuracy with higher levels of DSM participation. This occurs since higher DSM penetration results in more aggregated attacks. However, we note that an attacker, if having perfect knowledge of the grid could decrease their intensity and make detection more difficult. Furthermore, supervised learning classifiers performance on average was on par or better than sequential detection methods. LR detector had the highest accuracy for a lower level of DSM usage, while ANNs performed the worst in all scenarios. This highlights the power of linear detection methods over nonlinear. Point attacks resulted in the poorest detection with the CUSUM detector having the best performance. Type 2 attacks had higher recall and precision across all detectors for both levels of DSM. This is because a sudden attack shifts the mean of the data as a constant over time, which is more identifiable, yielding fewer false positives and more true positives.

\begin{table}[t]
	\centering
	\caption{Evaluation metrics for attack detection.}
	\label{t:metrics}
	\begin{tabular*}{\columnwidth}{@{\extracolsep{\fill}}rr|rrr|rrr@{}} 
		\toprule
		& $\kappa$ & \multicolumn{3}{ c| }{$0.1$} & \multicolumn{3}{ c }{$0.9$} \\
		\midrule
		& Attack Type	&	1	&	2	&	3	&	1	&	2	&	3	\\
		\midrule
		{\multirow{6}{*}{\rotatebox[origin=c]{90}{Accuracy}}} & LR	&	\textbf{89.6}	&	\textbf{100.0}	&	87.5	&	87.5	&	\textbf{100.0}	&	75.0	\\
		& RF	&	77.1	&	97.9	&	64.6	&	77.1	&	93.8	&	60.4	\\
		& GNB	&	70.8	&	77.1	&	75.0	&	79.2	&	\textbf{100.0}	&	72.9	\\
		& GBC	&	85.4	&	97.9	&	66.7	&	85.4	&	\textbf{100.0}	&	64.6	\\
		& ANN	&	56.3	&	95.8	&	20.8	&	75.0	&	79.2	&	14.6	\\
		& GLRT	&	79.2	&	95.8	&	58.3	&	\textbf{91.7}	&	97.9	&	58.3	\\
		& CUSUM	&	83.3	&	95.8	&	\textbf{95.8}	&	87.5	&	\textbf{100.0}	&	\textbf{100.0}	\\
		\midrule
		{\multirow{6}{*}{\rotatebox[origin=c]{90}{Recall}}} & LR	&	\textbf{92.0}	&	\textbf{100.0}	&	80.0	&	\textbf{89.0}	&	\textbf{100.0}	&	80.0	\\
		& RF	&	56.0	&	96.0	&	80.0	&	56.0	&	88.0	&	80.0	\\
		& GNB	&	88.0	&	56.0	&	80.0	&	60.0	&	\textbf{100.0}	&	80.0	\\
		& GBC	&	76.0	&	96.0	&	\textbf{100.0}	&	88.0	&	\textbf{100.0}	&	80.0	\\
		& ANN	&	68.0	&	92.0	&	60.0	&	76.0	&	68.0	&	80.0	\\
		& GLRT	&	76.0	&	92.0	&	80.0	&	84.0	&	96.0	&	80.0	\\
		& CUSUM	&	80.0	&	96.0	&	\textbf{100.0}	&	84.0	&	\textbf{100.0}	&	\textbf{100.0}	\\
		\midrule
		{\multirow{6}{*}{\rotatebox[origin=c]{90}{Precision}}} & LR	&	88.5	&	\textbf{100.0}	&	44.4	&	88.0	&	\textbf{100.0}	&	26.7	\\
		& RF	&	\textbf{100.0}	&	\textbf{100.0}	&	20.0	&	\textbf{100.0}	&	\textbf{100.0}	&	18.2	\\
		& GNB	&	66.7	&	\textbf{100.0}	&	26.7	&	\textbf{100.0}	&	\textbf{100.0}	&	25.0	\\
		& GBC	&	95.0	&	\textbf{100.0}	&	23.8	&	84.0	&	\textbf{100.0}	&	20.0	\\
		& ANN	&	56.7	&	\textbf{100.0}	&	7.7	&	76.0	&	89.5	&	9.1	\\
		& GLRT	&	82.6	&	\textbf{100.0}	&	17.4	&	\textbf{100.0}	&	\textbf{100.0}	&	17.4	\\
		& CUSUM	&	87.0	&	96.0	&	\textbf{71.4}	&	91.3	&	\textbf{100.0}	&	\textbf{100.0}	\\
		\bottomrule
	\end{tabular*}
\end{table}

\section{Conclusion} \label{s:future}

In this work, we study the exploitation of the hypothetical premise of the feedback between future cyber-enabled DSM programs, on the consumer side, and dynamic RTP on the utility side. An attacker with exploitive economic or nefarious intentions, such as causing efficiency loss of energy provision, can take advantage of the dependency between dynamic pricing and DSM load control. The utility modifies prices in response to forecasted demand in order to push realized load up or down. This is done to achieve some target load level with the goal to reduce peak load or achieve some other DSM objective. On the user side, cyber DSM programs then autonomously respond to prices to adjust certain portions of a users load up or down with some given elasticity. 

We propose two modes of attacks, false price data injections, and direct load manipulation. Under a false price data injection, an attacker modifies the RTP that users receive to alter their demand. Through a direct load manipulation attack, an attacker hacks and alters a users load profile directly. In both these attacks, aggregate load from the grid is modified which then can alter future prices or demand. We showcase how these two modes of attacks are equivalent and introduce three ways an attack can occur. The first type is a ramp attack, the second is a sudden attack, and the third is a point attack. We simulate these type of attacks and review several methods to detect them. 

We simulate and examine load-price data under different levels of DSM participation with three types of additive attacks: ramp, sudden, and point attacks. We applied sequential change point and supervised learning methods for detection of DSM attacks. Results conclude that higher amounts of DSM participation can exacerbate attacks but also lead to better detection of such attacks, point attacks are the hardest to detect, and supervised learning methods produce results on par or better than sequential detectors. Due to the speculative nature of these type of attacks, we plan to look into many different issues for future work. New directions include expanding our model to incorporate day ahead pricing and DSM scheduling, adding models of cost and utility on for a customer and ISO, expanding our model taking into account grid topology, and modeling advanced attack goals such as system failure by power line overload.

\newpage
\bibliographystyle{elsarticle-num}
\bibliography{mybib}

\end{document}